\documentclass{ar-astro2e} 
\usepackage{graphicx}
\usepackage{amssymb}
\usepackage{ARAstroBib} 
\begin{document}

%
\newcommand{\araa}{Annu. Rev. Astron. Astrophys.}
\newcommand{\apj}{Ap. J.}
\newcommand{\aj}{Astron. J.}
\newcommand{\apjs}{Ap. J. Suppl.}
\newcommand{\apjl}{Ap. J. L.}
\newcommand{\aap}{Astron. Astrophys.}
\newcommand{\aapr}{Astron. Astrophy. Rev.}
\newcommand{\aaps}{Astron. Astrophy. Suppl.}
\newcommand{\pasp}{Publ. Astron. Soc. Pac.}
\newcommand{\pasj}{Publ. Astron. Soc. Japan}
\newcommand{\physrep}{Physics Reports}
\newcommand{\mnras}{MNRAS}
\newcommand{\nat}{Nature}
\newcommand{\ssr}{Space Science Reviews}
\newcommand{\nar}{New Astronomy Reviews}
\newcommand{\herschel}{\emph{Herschel}}
\newcommand{\spitzer}{\emph{Spitzer}}
\newcommand{\planck}{\emph{Planck}}
\newcommand{\akari}{\emph{Akari}}
\newcommand{\iso}{ISO}
\newcommand{\alma}{ALMA}
\newcommand{\galex}{GALEX}
\newcommand{\wise}{WISE}
\newcommand{\arcmin}{\mbox{$^\prime$}}
\newcommand{\arcsec}{\mbox{$^{\prime \prime}$}}
\newcommand{\lsun}{\mbox{L$_\odot$}}
\newcommand{\msun}{\mbox{M$_\odot$}}
\newcommand{\msunyr}{\mbox{M$_\odot$~yr$^{-1}$}}
\newcommand{\rsun}{\mbox{R$_\odot$}}
\newcommand{\lbol}{$L_{\rm bol}$} 
\newcommand{\lfir}{$L_{\rm FIR}$} 
\newcommand{\lir}{$L_{\rm IR}$} 
\newcommand{\tdust}{$T_{\rm dust}$}
\newcommand{\mstellar}{$M_{\ast}$}     
\newcommand{\ergs}{erg~s$^{-1}$}    
\newcommand{\ergscm}{erg~s$^{-1}$~cm$^{-2}$} 
\newcommand{\mum}{\mbox{$\mu$m}}
\newcommand{\apss}{Ap\&SS}
\newcommand{\pasa}{PASA}

\newcommand{\Lnu}{\ifmmode L_{\rm \nu} \else $L_{\rm \nu}$\fi}
\newcommand{\Llambda}{\ifmmode L_{\rm \lambda} \else $L_{\rm \lambda}$\fi}
\newcommand{\LA}{\ifmmode L_{\rm 5100} \else $L_{\rm 5100}$\fi}
\newcommand{\Lg}{\ifmmode L_{\rm g(r)} \else $L_{\rm g(r)}$\fi}
\newcommand{\Ln}{\ifmmode L_{\rm n(r)} \else $L_{\rm n(r)}$\fi}
\newcommand{\Lo}{\ifmmode L_{\rm 0} \else $L_{\rm 0}$\fi}
\newcommand{\Ro}{\ifmmode R_{\rm 0} \else $R_{\rm 0}$\fi}
\newcommand{\Rhalf}{\ifmmode R_{\rm 1/2} \else $R_{\rm 1/2}$\fi}
\newcommand{\Rg}{\ifmmode R_{\rm g} \else $R_{\rm g}$\fi}
\newcommand{\Rms}{\ifmmode R_{\rm ms} \else $R_{\rm ms}$\fi}
\newcommand{\Rin}{\ifmmode R_{\rm in} \else $R_{\rm in}$\fi}
\newcommand{\Rout}{\ifmmode R_{\rm out} \else $R_{\rm out}$\fi}
\newcommand{\RISCO}{\ifmmode R_{\rm in} \else $R_{\rm in}$\fi}
\newcommand{\ro}{\ifmmode r_{\rm 0} \else $r_{\rm 0}$\fi}
\newcommand{\rhalf}{\ifmmode r_{\rm 1/2} \else $r_{\rm 1/2}$\fi}
\newcommand{\rg}{\ifmmode r_{\rm g} \else $r_{\rm g}$\fi}
\newcommand{\rms}{\ifmmode r_{\rm ms} \else $r_{\rm ms}$\fi}
\newcommand{\rin}{\ifmmode r_{\rm in} \else $r_{\rm in}$\fi}
\newcommand{\rout}{\ifmmode r_{\rm out} \else $r_{\rm out}$\fi}
\newcommand{\rISCO}{\ifmmode r_{\rm in} \else $r_{\rm in}$\fi}
\newcommand{\Fnu}{\ifmmode F_{\nu} \else $F_{\nu}$\fi}
\newcommand{\Flambda}{\ifmmode F_{\lambda} \else $F_{\lambda}$\fi}
\newcommand{\Mdot}{\ifmmode \dot{M\newcommand{\Lop}{\ifmmode L_{5100} \else $L_{5100}$\fi}} \else $\dot{M}$\fi}
\newcommand{\mdot}{\ifmmode \dot{m} \else $\dot{m}$\fi}
\newcommand{\Mrdot}{\ifmmode \dot{M}\left(r \right) \else $\dot{M}\left(r \right)$\fi}
\newcommand{\MRdot}{\ifmmode \dot{M}\left(R \right) \else $\dot{M}\left(R \right)$\fi}
\newcommand{\Mrindot}{\ifmmode \dot{M}\left(r_{ISCO} \right) \else $\dot{M}\left(r_{ISCO} \right)$\fi}
\newcommand{\Mroutdot}{\ifmmode \dot{M}\left(r_{out} \right) \else $\dot{M}\left(r_{out} \right)$\fi}
\newcommand{\MBHdot}{\ifmmode \dot{M}_{BH} \else $\dot{M}_{BH}$\fi}
\newcommand{\MBHdotexpct}{\ifmmode \dot{M}_{BHexpected} \else $\dot{M}_{BHexpected}$\fi}
\newcommand{\Medddot}{\ifmmode \dot{M}_{edd} \else $\dot{M}_{edd}$\fi}
\newcommand{\Moutdot}{\ifmmode \dot{M}_{out} \else $\dot{M}_{out}$\fi}
\newcommand{\Mindot}{\ifmmode \dot{M}_{in} \else $\dot{M}_{in}$\fi}
\newcommand{\Mwinddot}{\ifmmode \dot{M}_{wind} \else $\dot{M}_{wind}$\fi}
\newcommand{\MBH}{\ifmmode M_{\rm BH} \else $M_{\rm BH}$\fi}
\newcommand{\mbh}{\ifmmode M_{\rm BH} \else $M_{\rm BH}$\fi}
\newcommand{\Mexp}{\ifmmode M_{\rm 8} \else $M_{\rm 8}$\fi}

\newcommand{\Msun}{\ifmmode M_{\odot} \else $M_{\odot}$\fi}
\newcommand{\avisc}{\ifmmode \alpha_{visc} \else $\alpha_{visc}$\fi}

\newcommand{\Halpha}{\ifmmode {\rm H}\alpha \else H$\alpha$\fi}
\newcommand{\Hbeta}{\ifmmode {\rm H}\beta \else H$\beta$\fi}
\newcommand{\ka}{\ifmmode {\rm K}\alpha \else K$\alpha$\fi}
\newcommand{\Ka}{\ifmmode {\rm K}\alpha \else K$\alpha$\fi}
\newcommand{\ha}{\ifmmode {\rm H}\alpha \else H$\alpha$\fi}
\newcommand{\hb}{\ifmmode {\rm H}\beta \else H$\beta$\fi}
\newcommand{\MgII}{\ifmmode {\rm Mg}\,\textsc{ii}\,\lambda2798 \else Mg\,{\sc ii}\,$\lambda2798$\fi}
\newcommand{\mgii}{\ifmmode {\rm Mg}{\textsc{ii}} \else Mg\,{\sc ii}\fi}
\newcommand{\CIV}{\ifmmode {\rm C}\,\textsc{iv}\,\lambda1549 \else C\,{\sc iv}\,$\lambda1549$\fi}
\newcommand{\civ}{\ifmmode {\rm C}\,\textsc{iv} \else C\,{\sc iv}\fi}
\newcommand{\feii}{Fe\,{\sc ii}}
\newcommand{\feiii}{Fe\,{\sc iii}}
\newcommand{\oi}{\ifmmode \left[{\rm O}\,\textsc{i}\right] \else [O\,{\sc i}]\fi}
\newcommand{\OI}{\ifmmode \left[{\rm O}\,\textsc{i}\right]\,\lambda6300 \else [O\,{\sc i}]$\,\lambda6300$ \fi}

\newcommand{\nii}{\ifmmode \left[{\rm N}\,\textsc{ii}\right] \else [N\,{\sc ii}]\fi}
\newcommand{\NII}{\ifmmode \left[{\rm N}\,\textsc{ii}\right]\, \lambda6584 \else [N\,{\sc ii}]\,$\lambda6584$\fi}
\newcommand{\SII}{\ifmmode \left[{\rm S}\,\textsc{ii}\right]\, \lambda6731 \else [S\,{\sc ii}]\,$\lambda6731$\fi}
\newcommand{\oii}{\ifmmode \left[{\rm O}\,\textsc{ii}\right] \else [O\,{\sc ii}]\fi}
\newcommand{\OII}{\ifmmode \left[{\rm O}\,\textsc{ii}\right]\,\lambda3727 \else [O\,{\sc ii}]\,$\lambda3727$ \fi}
\newcommand{\oiii}{\ifmmode \left[{\rm O}\,\textsc{iii}\right] \else [O\,{\sc iii}]\fi}
\newcommand{\OIII}{\ifmmode \left[{\rm O}\,\textsc{iii}\right]\,\lambda5007 \else [O\,{\sc iii}]\,$\lambda5007$\fi}
\newcommand{\oiv}{\ifmmode \left[{\rm O}\,\textsc{iv}\right] \else [O\,{\sc iv}]\fi}
\newcommand{\ov} {\ifmmode \left[{\rm O}\,\textsc{v}\right] \else [O\,{\sc v}]\fi}
\newcommand{\ovi}{\ifmmode \left[{\rm O}\,\textsc{vi}\right] \else [O\,{\sc vi}]\fi}
\newcommand{\ovii}{\ifmmode \left[{\rm O}\,\textsc{vii}\right] \else [O\,{\sc vii}]\fi}
\newcommand{\sivi}{\ifmmode \left[{\rm Si}\,\textsc{vi}\right] \else [Si\,{\sc vi}]\fi}
\newcommand{\OIV}{\ifmmode \left[{\rm O}\,\textsc{iv}\right]\,\lambda25.9\mu m \else [O\,{\sc iv}]\,$\lambda25.9\mu m$\fi}
\newcommand{\OVII}{\ifmmode {\rm O}\,\textsc{vii}\,570 eV \else O\,{\sc vii}\,$570\,eV$\fi}
\newcommand{\SiVI}{\ifmmode \left[{\rm Si}\,\textsc{vi}\right]\,\lambda1.96\mu m \else [Si\,{\sc vi}]\,$\lambda1.96\mu m$\fi}
\newcommand{\NeV}{\ifmmode \left[{\rm Ne}\,\textsc{v}\right]\,\lambda3426 \else [Ne\,{\sc v}]\,$\lambda3426$\fi}

\newcommand{\HeIIop}{He\,{\sc ii}\,$\lambda4686$}
\newcommand{\HeIIuv}{He\,{\sc ii}\,$\lambda1640$}
\newcommand{\OIIIuv}{O\,{\sc iii}]\,$\lambda1663$}
\newcommand{\NIVuv}{N\,{\sc iv}]\,$\lambda1486$}

\newcommand{\ld}{\ifmmode {\rm lt-days} \else lt-days \fi}
\newcommand{\kms}{\ifmmode {\rm km\,s}^{-1} \else km\,s$^{-1}$ \fi}
\newcommand{\cc}{\hbox{cm$^{-3}$}}
\newcommand{\cmii}{\hbox{cm$^{-2}$}}
\newcommand{\ergcmsA}{\ifmmode{\rm erg}\, {\rm cm}^{-2}\,{\rm s}^{-1}\,{\rm\AA}^{-1} \else erg\, cm$^{-2}$\, s$^{-1}$\, \AA$^{-1}$\fi}
\newcommand{\ergcmsHz}{\ifmmode{\rm erg\,cm}^{-2}\,{\rm s}^{-1}\,{\rm Hz}^{-1} \else erg\,cm$^{-2}$\,s$^{-1}$\,Hz$^{-1}$\fi}
\newcommand{\phcms}{\ifmmode {\rm ph\,cm}^{-2}\,{\rm s}^{-1} \else ,ph\,cm$^{-2}$\,s$^{-1}$\fi}
\newcommand{\phcmsA}{\ifmmode {\rm ph\,cm}^{-2}\,{\rm s}^{-1}\,{\rm\AA}^{-1} \else ph\,cm$^{-2}$\,s$^{-1}$\,\AA$^{-1}$\fi}

\newcommand{\Lsun}{\ifmmode L_{\odot} \else $L_{\odot}$\fi}
\newcommand{\auvo}{\ifmmode \alpha_{\nu,{\rm UVO}} \else $\alpha_{\nu,{\rm UVO}}$\fi}
\newcommand{\Luv}{\ifmmode L_{1450} \else $L_{1450}$\fi}
\newcommand{\Lop}{\ifmmode L_{5100} \else $L_{5100}$\fi}
\newcommand{\Lhree}{\ifmmode L_{3000} \else $L_{3000}$\fi}
\newcommand{\lLthree}{\ifmmode \log\left(\Lthree/\ergs\right) \else $\log\left(\Lthree/\ergs\right)$\fi}
\newcommand{\lledd}{\ifmmode L/L_{\rm Edd} \else $L/L_{\rm Edd}$\fi}
\newcommand{\Ledd}{\ifmmode L_{\rm AGN}/L_{\rm Edd} \else $L_{\rm AGN}/L_{\rm Edd}$\fi}
\newcommand{\lamLlam}{\ifmmode \lambda L_{\lambda} \else $\lambda L_{\lambda}$\fi}
\newcommand{\Lbol}{\ifmmode {\rm L}_{\rm bol} \else L$_{\rm bol}$\fi}
\newcommand{\lLbol}{\ifmmode \log\left(\Lbol/\ergs\right) \else $\log\left(\Lbol/\ergs\right)$\fi}
\newcommand{\Tdust}{\ifmmode T_{\rm dust} \else $T_{\rm dust}$\fi}

\newcommand{\Fthree}{\ifmmode F_{3000} \else $F_{3000}$\fi}
\newcommand{\mic}{$\mu {\rm m}$}

\def \XMM{{\it XMM-Newton}}
\def \Chandra{{\it Chandra}}
\def \chandra{{\it Chandra}}
\def \herschel{{\it Herschel}}
\def \Herschel{{\it Herschel}}
\def \spitzer {{\it Spitzer}}
\def \Spitzer {{\it Spitzer}}
\def \wise  {{\it WISE}}
\def \WISE  {{\it WISE}}
\def \iso  {{\it ISO}}
\def \ISO  {{\it ISO}}
\def \akari  {{\it AKARI}}
\def \AKARI  {{\it AKARI}}
\def \SWIFT  {{\it SWIFT}}
\def \swift  {{\it SWIFT}}
\def \INTEGRAL  {{\it INTEGRAL}}
\def \integral  {{\it INTEGRAL}}
\def \LAGN {L$_{\rm AGN}$}
\def \Ldust {L$_{\rm dust}$}
\def \LDUST {L$_{\rm dust}$}
\def \LSF {L$_{\rm SF}$}
\def \Lbol {L$_{\rm bol}$}
\def \Cf {$f_C$}
\def \cf {$f_C$}
\def \CF {$f_C$}
\def\cblue{\color{blue}}
\def\cred{\color{red}\sout}
\jname{Annu. Rev. Astron. Astrophys.}
\jyear{2015}
\jvol{53}
\ARinfo{--------}

\title{Revisiting the Unified Model of Active Galactic Nuclei}

\markboth{Netzer}{AGN Unification}

\author{Hagai Netzer
\affiliation{School of Physics and Astronomy, Tel Aviv University, Tel Aviv 69978, Israel, email:netzer@wise.tau.ac.il}}


\begin{abstract}
This review describes recent developments related to the unified model of active galactic nuclei (AGN).
It focuses on new ideas about the origin and properties of the central obscurer (torus), and the connection 
with its surrounding. The review does not address radio unification. 
AGN tori must be clumpy 
but the uncertainties about their properties are still large. Todays most promising models involve 
disk winds of various types and hydrodynamical simulations that link the large scale galactic disk to the inner 
accretion flow. IR studies greatly improved the understanding of the spectral 
energy distribution of AGNs but they are hindered by various selection effects. X-ray samples are more complete.  
A basic relationship which is still unexplained is the dependence of the torus  
covering factor on luminosity. There is also much confusion regarding ``real type-II AGNs'' that do not 
fit into a simple unification scheme. The most impressive recent results are due to 
IR interferometry,
which is not in accord with most torus models, and the accurate mapping of central ionization cones.
AGN unification may not apply to merging systems and is possibly restricted to secularly evolving galaxies.

\end{abstract}

\maketitle


\section{THE AGN FAMILY: CLASSIFICATION AND UNIFICATION}
\label{sec:the_agn_family}

\subsection{AGN Classification}
\label{sec:classification}

An active galactic nucleus (AGN) is defined here as a galaxy containing a massive ($>10^5$ \msun) accreting black hole (BH) with
Eddington ratio exceeding the (somewhat arbitrary) limit of \Ledd$= 10^{-5}$, where \LAGN\ is the bolometric luminosity and 
L$_{Edd}=1.5 \times 10^{38} M_{BH}$/\msun\ \ergs\
is the Eddington luminosity for a solar composition gas. 
This definition excludes the milky way Galaxy but includes a large
number of low ionization nuclear emission-line regions (LINERs) and low ionization radio galaxies. 
Most AGNs include several of the following components:
\begin{itemize}
 \item
 A sub-pc rotational dominated accretion flow that is usually refers to as an accretion disk.
Optically thick disks can be geometrically thin (``thin'' accretion disk) or thick
(``slim'' or ``thick'' accretion disk\footnote{
The terminology here is not very clear and both names are used for systems that are not geometrically thin. In this
article I use the term ``slim disk''.}).
 Optically thin accretion disks or flows that are occasionally advection dominated. Such structures are referred to as  
radiation inefficient accretion flow (RIAF), or advection dominated accretion flow (ADAF) \citep{Narayan2004}.
 \item
 High density, dust-free gas clouds
moving at roughly Keplerian velocities at a luminosity dependent distance of 0.01-1 pc from the BH (the broad line region, BLR)
\item
An axisymmetric dusty structure with luminosity dependent dimensions of 0.1--10 pc (the central ``torus'').
\item
Lower density, lower velocity ionized gas (narrow line region, NLR) extending from just outside the torus to hundreds and 
even thousand of parsecs along
the general direction of the opening in the torus (``ionization cones''). Most of this gas contains dust except for very close in, in a region 
referred to as ``the coronal line region''.
\item
A very thin molecular maser disk with size similar to that of the torus.
\item
A central radio jet occasionally associated with $\gamma$-ray emission..
\end{itemize}
These components have been discussed, extensively, in numerous books and review articles 
\citep{Robson1996,Peterson1997,Krolik1998,Blandford1990,Osterbrock2006,Netzer2013,Antonucci1993,Ho2008,Heckman2014} and the reader is referred to these
 references for more information. 

``AGN unification'' is the idea that the large diversity of observed AGN properties can be explained by a small number of physical parameters. 
The old unification scheme \citep{Antonucci1993,Urry1995,Urry2003}
is a courageous attempt to combine the ever increasing number of sub-groups into a general picture with two parameters:
the torus inclination to the line of sight (LOS) and the source luminosity (``unification by inclination''). This scheme is perhaps
the simplest possible way to characterize the known fact that the nuclear continuum and emission line radiation of AGNs
 can suffer wavelength dependent
scattering, absorption and
reflection on the way out. This can take place in the torus, in the disk of the host galaxy, in stellar and nuclear
outflows, and inside the BLR itself. 
Recently it was suggested that AGNs can be separated in a different way into two 
major groups: ``radiative mode'' and ``jet mode'' AGNs \citep{Heckman2014}.
Most of the energy output in radiative-mode AGNs is in the form of electromagnetic radiation, and is a direct result
of matter accretion through a central optically thick accretion disk. This group is referred to in the literature
as Seyfert galaxies or QSOs. About 10\% of the sources in this group are radio-loud, showing a highly collimated, relativistic radio jet
and, occasionally, a $\gamma$-ray jet. 
Radiative mode AGNs are very efficient accretors with \Ledd$\geq 0.01$. An alternative name, based on the level of ionization of the NLR gas, 
is ``high ionization AGNs'' (referred to also as ``high excitation AGNs'').
The prime energy output of jet-mode AGNs is bulk kinetic energy transported
in two-sided jets. Their typical Eddington ratio is much smaller and the jets are most likely powered via a RIAF.
The members of this group are low luminosity radio galaxies and LINERs, and an alternative spectroscopic definition is ``low ionization'' AGNs.
In the local universe the mean BH mass of objects in the second group is larger. 

This review concerns mostly radiative mode AGNs and the reader is referred to \citet{Ho2008} and \citet{Heckman2014} for more information
about radio mode sources. The exception are some LINERs with total electromagnetic radiation that equals and even exceeds, the jet kinetic energy. 
I will not discuss radio-loud, radiative mode AGNs although most of their properties are indistinguishable from those
of the radio-quiet sources. This excludes about 10\% of radiative mode AGNs.

The main sub-groups addressed in this review are: \\
\noindent
{\bf Type-I AGNs:}
Sources showing broad ($1000-20,000$ km~sec$^{-1}$) permitted and semi-forbidden emission lines and a bright, non-stellar, central point source visible
at all wavelengths that are not contaminated by stellar light.
Almost all low to intermediate luminosity type-I AGNs show strong, high ionization narrow emission lines many of which are forbidden lines.
Most papers refer to these sources as Seyfert 1 galaxies, or QSOs.
Narrow emission lines are missing from the spectrum of many high luminosity
type-I AGNs. Many papers make a distinction between several sub-groups of
type-I AGNs like type 1.5, type 1.8 or type 1.9. This refers to the relative intensity of the broad and narrow
components of the Balmer lines, type 1.9 being the ones with the stronger narrow component. Here I refer to all such objects as 
type-Ii where ``i'' stands for intermediate (in some papers ``i'' refers to broad Paschen lines).
This has been a source of much confusion
since part of the spectroscopic characteristics have little to do with the conditions near the BH (e.g. reddening in the host galaxy)
and weak broad lines can also be the result of line and continuum variations.

\noindent
{\bf Type-II AGNs:}
 Sources containing strong narrow ($300-1000$ km~sec$^{-1}$) NIR-optical-UV emission lines that are somewhat broader than what
is observed in emission line galaxies of similar types. The lines show clear indications of photoionization
by a non-stellar source. The strongest lines are \OIII, \NII, \OII, \OIV, \NeV, \CIV\ and the hydrogen Balmer and Lyman lines.
Like type-I AGNs, they also show a point X-ray source. 
Advances in mid-IR (MIR) spectroscopy resulted in attempts to identify type-II AGNs by the equivalent width (EW) of the
silicate 9.7\mic\ absorption feature in their spectrum. Such methods are not in very good agreement with optical classification.
Type-II AGNs are further divided into two subgroups. The first contains hidden type-I sources with broad emission lines
seen in polarized light. 
The second group is less accurately defined and is occasionally referred to as ``real type-II'' or ``true type-II'' AGNs. 
They show similar width and excitation narrow lines but no detectable 
broad lines and little X-ray absorption. Their mean luminosity is below the luminosity of the type-II objects with hidden broad lines.
In the local Universe, such objects represent about 30\% of all type-II objects 
\citep[][]{Panessa2002,Tran2003,Shi2010,Brightman2011b,Marinucci2012,Merloni2014}.

\noindent
{\bf LINERs:}
AGNs that are characterised by their low ionization, narrow emission lines from gas which is 
ionized by a non-stellar source. Typical strong emission lines in this group are \NII, \NII\ and \SII, and the Balmer lines.
I will only address LINERs with EW(\ha)$\geq 3$\AA\ to avoid the confusion with gas ionized by post-AGB stars \citep[e.g.][]{CidFernandes2011}. 
LINERs can be divided into type-I (broad emission lines) and type-II (only narrow emission lines) LINERs.
Some but not all LINERs show point-like X-ray and UV sources and
UV and X-ray variations \citep{Maoz2007,Hernandez2013}. Detailed reviews of LINERs are given in \citet{Ho2008} and \citet{Heckman2014}.

\noindent
{\bf Lineless AGNs:}
This subgroup consists of AGNs with extremely weak, sometimes completely undetected emission lines. They are recognized by the 
presence of a non-stellar central source and, occasionally, continuum variability.
This general group includes two distinct sub-groups: low luminosity sources \citep[e.g.][]{Trump2009}, and very
high luminosity sources \citep[e.g.][]{Shemmer2010,Meusinger2014}. 

Classifying type-I AGNs is relatively easy except for cases where the BLR is heavily reddened by foreground dust. 
Identifying and classifying narrow line AGNs is more challenging, especially when no hard X-ray data are available, partly because
pure star forming (SF) galaxies (also referred to as HII galaxies), that are more numerous than AGNs, also show the same strong narrow emission lines.
An efficient way to separate the groups is to make use of 
``diagnostic diagrams'' 
\citep[see][]{Baldwin1981,Veilleux1987,Kauffmann2003,Kewley2006} that compare various line ratios that differ according to the nature of the ionizing
continuum (stellar or non-stellar). The most useful line ratios are \OIII/\hb, \NII/\ha, and \OI/\ha. 
Diagnostic diagrams show several well defined regions, one including high ionization AGNs, one including mostly LINERs, and one occupied by 
SF galaxies. The region between the pure SF systems and AGNs is occupied by ``composite sources'', with intermediate spectral properties.

Most AGNs known to date were found in large surveys, like the Sloan Digital Sky Survey (SDSS), through spectroscopic observations with relatively
large apertures (e.g. 3 arc-sec  in the SDSS). 
The host galaxies of most AGNs are SF galaxies \citep{Rosario2012,Rosario2013}. We can estimate the total emitted SF luminosity, \LSF,
and distinguish  ``AGN dominated'' (\LAGN$>$\LSF) from ``SF dominated'' (\LSF$>$\LAGN) systems. For example, in the spectroscopic 3 arc-sec SDSS
sample, such an comparison 
shows that many composite sources are SF dominated galaxies. 
Difficulties in identifying AGNs in composite sources can affect the census of local type-II AGNs, their luminosity and mass functions (LFs, MFs). 
The census of type-I AGNs is  based on other methods and is not subjected to such uncertainties beyond a redshift of about 0.1.

\begin{table}[th!]
 \begin{tabular}{lcccc}
 Redshift  & SF galaxies & LINERs  & Type-II AGNs & Composite sources \\
 0.05--0.1 & 0.755        & 0.035   &   0.04      &     0.17         \\
 0.15--0.2 & 0.605       & 0.035   &   0.11      &     0.25           \\
 \end{tabular}
 \caption{The fraction of strong emission line sources, at two redshift intervals,
   as derived from diagnostic diagrams of SDSS objects from the data release 7 of the catalog.
   The total number of sources in the 0.05--0.1 redshift interval is 108,353 and the number in
the 0.15--0.2 interval, 17,572.
  The selection includes all emission line objects where the lines that were used for the classification  have
 signal-to-noise ratio larger than 3. This introduces a large bias against sources at higher redshifts, especially weak
line LINERs (courtesy of David Rosario).
 }
\end{table}

\subsection{The Observed Signature of a Central Obscurer}
\label{sec:central_obscurer}

\subsubsection{Optical-UV Absorption and Reflection}

The original unification scheme \cite[][and references therein]{Antonucci1993} included a central axisymmetric dusty structure with a column density
which is large enough to completely obscure the central source in some directions. This is usually referred to as the central ``torus'';
a somewhat ambiguous name which will, nevertheless, be retained here. The simplest torus is made of
a smooth matter distribution. More elaborated structures, made of clumps and inter-clump material, are preferred by today's observations.
The gas at the inner radius of the torus is ionized by the central source. Deeper in, the torus contains dusty molecular gas. 

It is interesting to study the observational consequences of placing such a simple geometrical structure around the central BH and accretion disk.
The first prediction is a dependence of the obscuring column on inclination to the LOS, where inclination is measured relative to 
the $z$ axis of the system (the direction perpendicular to the central disk).
Small inclinations, close to face-on, allow a clear view of the central source
and high inclinations result in large obscuration at UV-optical wavelengths.

The second  prediction is based on the fact that electromagnetic 
radiation reflected off dust grains and free electrons is polarized, with angle and percentage polarization which depend on  
geometry and wavelength. 
The seminal work of \cite{Antonucci1985} is among the first of many papers showing how to detect 
hidden broad emission lines using spectropolarimetry
and how to infer the BLR kinematics and overall geometry from such observations.
Later and deeper polarimetry \cite[e.g.][]{Tran2003} show a large fraction of type-IIs with no hint for polarized broad lines. This
led to the suggestion 
that objects lacking polarized broad lines are unobscured but void of BLR gas, like the real type-II AGNs (\S~\ref{sec:classification})
that show no detectable broad lines even when not obscured..

\subsubsection{X-ray Absorption Reflection and Emission} 

The next unavoidable consequence is a column density dependent X-ray absorption. The assumption used here is of
an isotropically emitting central X-ray source, possibly a hot corona over the central part of the accretion disk.
For a Compton thin torus, with $N_H<1.25 \times 10^{24}$ cm$^{-2}$, the obscuring column density can be measured by modeling the observed
spectrum provided the intrinsic spectral energy distribution (SED) 
and the gas metallicity are known (for solar metallicity gas, most of the X-ray absorption beyond 0.3 keV is due to metals).

There are several X-ray spectral features that can help diagnose the torus properties.
The first is the 
``Compton hump'', an excess radiation at around 20 keV due to the down scattering 
of hard X-ray photons. Almost all type-I AGNs show Compton humps in their spectra
(the terms ``scattering'' and ``reflection'' are used quite loosely in the literature, all referring
to Compton scattering followed by absorption, see e.g. \cite{Netzer1993a}).

The second feature is due to reflection off electrons in highly ionized gas.
Here the reflected radiation is a mirror image of the intrinsic continuum with intensity that, for all practical
purposes, depends only on the geometrical covering factor (\cf=$\Omega/4 \pi$ where $\Omega$ is the solid angle subtended by the torus). 
This fraction hardly ever exceeds a few percent.

The third spectral feature
is the K$_{\alpha}$ iron line. This emission feature is a collection of fluorescence (Auger) K$_{\alpha}$ transitions and recombination of
H-like and He-like iron at energies between 6.4 and 7 keV,
depending on the specific ion. The lines are broad and of high ionization
 if originating from the surface of the central accretion disk, or narrow and occasionally of low ionization if due to X-ray illumination of the cold gas in the torus.
 In case where the X-ray continuum is unobscured, the EW of the narrow \Ka\ line depends on the number of absorbed E$>7.1$ keV photons. 
For a typical X-ray SED and solar metallicity gas, this is estimated to be: 
EW(\Ka)$\approx$\Cf$\times N_H / 10 ^{21}$ eV,
where complications like multiple scattering have been ignored 
 (note that $N_H$ is not necessarily the LOS absorbing column, see \S3.3)
For a torus with a covering factor of 0.5 and column density of $10^{23}$ cm$^{-2}$, EW(\Ka)$\sim 50-100$ eV.
Indeed, many type-I AGNs show narrow \Ka\ lines with this EW \cite[][and references therein]{Bianchi2008,Ricci2013}.
An even clearer signature of a central torus is the very large EW(\ka) in Compton thick sources.
In this case, much of the \ka\ line reaches the observer without being absorbed but most of the central
continuum radiation is blocked by the torus. The observer can only see the much weaker reflected continuum which
raises the observed EW dramatically and it can reach 1--2 keV. 

\subsubsection{Ionization Cones}

The unobscured ionizing radiation of the central source reaches the gas in the galaxy causing heating and ionization.
This results in the emission of various narrow lines.
A torus like geometry will result in two ``ionization cones'' (or bicone)
 Depending on the geometry and inclination to the line of sight, one of the cones may not be observed.

\subsubsection{The Dusty Boundary of the Broad Line Region} 
The dust sublimation temperature, $T_{\rm sub}$, is the maximum
temperature attained by a dust grain before it evaporates. This depends on grain size and composition and the local flux.
 For ISM type graphite grains, 
the maximum sublimation temperature is 
$T_{\rm sub} \simeq 1800$\,K and for silicate grains,
$T_{\rm sub} \simeq 1400$\,K.
These numbers can be used to derive a  {\it mean sublimation radius}, $\langle R_{\rm sub} \rangle $,
which is the minimum radius where a grain of a certain size can survive the local radiation field. 
Averaging over ISM-type grain sizes, one obtains the following mean sublimation distances for graphite (C) 
and silicate (Si) grains \cite[e.g.][]{Barvainis1987}:
\begin{equation}
\langle R_{\rm sub,C} \rangle \simeq 0.5 L_{46}^{1/2}  \left [ \frac{1800}{T_{\rm sub}} \right ]^{2.6} f(\theta) \,\, {\rm pc}\, ;  
\,\,\,\,\,\,\,
%
\langle R_{\rm sub,Si} \rangle \simeq 1.3 L_{46}^{1/2} \left [ \frac{1500}{T_{\rm sub}} \right ]^{2.6} f(\theta) \,\, {\rm pc} \, , 
\label{eq:r_sub}
\end{equation}
where $L_{46}=$\LAGN/$10^{46}$ \ergs\ and $f(\theta)$ is an angular dependent term that allows for anisotropy of the central 
source radiation \citep{Netzer2014a}.
For an isotropic source 
$f(\theta)=1$ and for a thin disk with electron scattering atmosphere, 
$f(\theta)= \left [ \cos \theta (1+2 \cos \theta) / 3 \right]^{1/2}$ \citep{Netzer2013}. 
The above numbers are computed for a mean grain size of $a=0.05$\mic\ and the distance depends roughly on $a^{-2}$.

Dust can compete effectively with the ionization of the gas
because of its large absorption cross sections at all wavelengths longer than about
$0.02\, \mu$m. For highly ionized gas, 
$  N_{\rm dust} /  N_{\rm H^0}  \propto N_{\rm gas}/N_{\rm H^0}
               \propto U_{\rm H}$,
where $U_{\rm H}$ is the 
ionization parameter defined over the hydrogen Lyman continuum \citep{Netzer2013}.
At high levels of ionization, the dust absorbs a larger fraction of the photons capable
of ionizing hydrogen and helium. This reduces the size of the ionized part of the cloud and lowers the intensity of most 
emission lines. 
Emission line photons produced far from the surface of an ionized, dusty gas cloud, suffer additional attenuation on their way out due to
absorption by the dust. This is most noticeable in lines with high optical depth where the large number of scatterings of the line photons increase considerably
the path length.
Thus, a clear prediction of the torus scenario is that the high velocity gas near the BH will be divided into two distinct regions, dust-free
gas with strong line emission and dusty gas with weak line emission \citep{Netzer1993}.

\subsubsection{Infrared Emission}

A clear prediction of any torus model is a broad IR SED with a total luminosity that depends almost entirely 
on the fraction of the central source radiation absorbed by dust in the torus (a combination of optical depth and covering factor).
The radiation may be emitted isotropically, or anisotropically, depending
on the optical depth and exact geometry. The SED is predicted to show several silicate 
features centered around 9.7 and 18\mic, either in absorption or emission, depending on the exact geometry and LOS optical depth.
The emitted radiation of basically all AGNs is varying in time.
This will result in a time-dependent incident flux on the inner torus walls, and
time dependent NIR and MIR emission by the dust.

\subsection{Unification Schemes}

The term ``unification scheme'' refers usually to two different categories: IR-optical-UV-X-ray unification, and radio unification.
The first involves a central dusty torus and can explain the major differences between type-I and type-II AGNs with a 
surprisingly small number of assumptions. 
This was first suggested in the late 1980s. The second combines the torus with a 
relativistic jet observed in some 10\% of all high ionization AGNs. The jet is launched in the vicinity of the BH and is aligned with
the symmetry axis of the system. This can be used to make specific predictions about the observed properties of compact and extended radio-loud AGNs. 
Radio unification is not covered in this review and the reader is referred to \cite{Urry2003} and \cite{Tadhunter2008} 
for references on this topic.
It is also worth noting that LINERs were not included in the first unification schemes.

Recent observations of large AGN samples, and more detailed information about specific sources, lead to the
conclusion that the old unification scheme requires three major modifications.
The first is related to the physics and structure of the central obscurer.
The second is the realization that there are real differences between various sub-groups, beyond the dependence on luminosity and torus inclination. 
This can be due to the nature of the central power-house (e.g. a RIAF instead of high efficiency accretion flow), 
and the fact that some AGNs lack one or more of the main components, e.g. a BLR or a NLR. The third is related to BH and galaxy evolution.
The present review focuses on these modifications.

\subsection{The Era of Selection Effects}

Understanding the cosmological evolution of super-massive BHs is key to the understanding of galaxy evolution 
This requires reliable luminosity and mass functions which depend on AGN classification. While large surveys like SDSS contain hundreds of thousands
of AGNs up to $z \sim 7$, this survey, and basically all others, are hindered by various selection effects.
Since the topic of this review is AGN unification, I will only mention here selection effects that can bias AGN classification.

The use of diagnostic diagrams in spectroscopic surveys, like SDSS, to obtain complete AGN samples is limited to low redshift.
Beyond $z \sim 0.2$, type-II AGNs with weaker emission lines drop from the sample because of poor signal to noise (S/N) and contamination
by stellar absorption lines (weak line LINERs are already missing from such samples at redshift as small as 0.1).
At $z\geq 0.4$, the \ha\ line, which is crucial for the classification, is shifted beyond the observable range. An alternative method to detect
luminous type-II AGNs up to $z \sim 0.83$ \cite[][and refferences therein] {Reyes2008} is to search for strong \oiii\ lines relative to \hb\
which is required {\it not to show} a broad component. Unfortunately, 
large IR spectroscopic surveys are not yet available, and alternative diagnostics, based on emission lines at shorter wavelengths, are not as 
reliable. Broad line AGNs are detected by their continuum shape and their broad emission lines. At $z\leq 0.1$, the least luminous
type-Is drop from such samples due to strong stellar continuum and large amplitude continuum variations.

Reddening and obscuration is another reason for losing sources from large samples. At small redshift, this affects type-IIs more than type-Is. 
IR-selected samples are less prone to reddening but are biased towards sources with larger covering factor and against large inclination
angles (see \S\ref{sec:torus_models}). They are also contaminated by emission from
SF regions in the galaxy. Mid-IR photometric surveys, like those based on \spitzer\ \citep{Stern2005} are very efficient in differentiating AGNs
from non-active galaxies. However, as will be shown later, the IR-SED is not
uniform enough to unambiguously differentiate between AGN types and type-II classification is most problematic. 
Host galaxy contamination affects the measuring of low luminosity AGNs over the entire rest-frame range of 0.3--2 \mic.
Selecting and classifying AGNs by their hard X-ray continuum is, arguably, the most reliable way at almost all redshifts. 
The completeness of this method depends on the energy and flux limit of the specific sample
and various sophisticated methods have been developed to allow for such biases. A major difficulty is 
to detect highly obscured, large column density  X-ray sources at high redshift.
 
A comprehensive comparison of the efficiency of various methods to discover AGNs is given in \citet{Merloni2013}. 
According to this work (see their Figure 16 and related explanation), the missing fractions of AGNs in various samples depend on the bolometric luminosity and
are roughly:
  Optical (B-band) samples, from 70\% at \LAGN=$10^{47}$ erg~sec$^{-1}$ to 90\%  at \LAGN=$10^{42}$ erg~sec$^{-1}$.
  MIR (15\mic) samples, from 50\% at \LAGN=$10^{47}$ erg~sec$^{-1}$ to 70\%  at \LAGN=$10^{42}$ erg~sec$^{-1}$.
  Hard X-ray  (2--10 keV) samples, from 0\% at \LAGN=$10^{47}$ erg~sec$^{-1}$ to 40\%  at \LAGN=$10^{42}$ erg~sec$^{-1}$.

The following sections review todays torus models (\S2), and torus observations (\S3). This is followed by a description of additional, 
smaller scale and larger scale obscurers (\S4),
ionization cones (\S5) and other galactic components connected to the torus (\S6).
Finally, in \S7, I explain why AGN unification may not apply to major mergers. \S8 is a short summary of the main points of this review.

\nopagebreak
\section{DISK AND TORUS MODELS}
\label{sec:torus_models}

\subsection{Central Disks and Central Tori}

How large and how massive is the central structure? How is it connected to the BLR and accretion disk on its inner side, and to the ISM
of the host galaxy on the outer side? Is it a torus, a disk, or some other shape? Is such a structure typical of all large bulge galaxies or is BH accretion
essential to form and maintain it? 
And most important, how is this structure formed in active galaxies?
 The global theoretical framework is presented in this section and the observational consequences in \S3.

The terminology used in this area of research requires some clarification since the terms ``central disk'', ``central torus'' and ``central accretion disk'' 
have been used
quite loosely. In this review, the term inner accretion disk refers to the accretion flow inside the sublimation radius, where the accreted gas is losing
angular momentum through molecular, magnetic or other sources of viscosity. Many AGNs must contain a much larger
disk-like structure, 1-100 pc in size, that connects the inner disk to the host galaxy. The inner part of this structure can have
a thick geometry in which case it becomes the inner torus. The entire structure, that can extend beyond the torus, is referred
to here as the ``central disk'' or the ``$Q$ disk'' (after the Toomre stability parameter $Q$, see below). It contains a large amount of molecular gas that, 
given a large enough column density, undergoes SF associated with super-novae (SN) explosions, the growth of a nuclear star cluster, etc. 
Some of these processes result in local effective viscosity that can make this entire structure a large accretion disk, where gas loses its angular momentum and drifts 
into the center. The angular momentum of the large disk is not necessarily aligned with the angular
momentum of the inner disk. There is also no known reason why such a large disk will maintain its properties
throughout the accretion episode.

\subsection{General Models}

Two general types of central disks and central tori have been considered. The first are theoretical models 
\citep[e.g.][]{Krolik1988,Collin1999,Emmering1992,Schartmann2005,Collin2008,Beckert2004,Honig2007,Krolik2007,Elitzur2006,Vollmer2008,Wada2009,Wada2012,Schartmann2010}
that were developed to explore specific aspects like the formation, stability, shape and evolution of the central structure. 
The models attempt to answer two fundamental questions: what is the mechanism that removes angular momentum from galactic infalling gas thus allowing the feeding of
the central BH, and what are the ways to maintain a thick gaseous structure for a long period of time.
Important processes considered in this context include blob motions along magnetic field lines, colliding magnetic clumps, 
radiation pressure supported dusty clumps, SF driven turbulence, SN explosions, AGN and stellar feedback via radiation pressure and winds, and more. 
Several analytic and semi-analytic models have been proposed 
but the hydrodynamical simulations seem to provide more realistic ways to model this complex situation.
The second type is phenomenological, ad-hoc models that make assumptions about the size, composition  and geometry of the torus in an attempt to explain the 
observations, mostly the spatially resolved images and the NIR-MIR SED 
\cite[e.g.][]{Pier1992,Fritz2006,Nenkova2008a,Honig2010,Stalevski2012}. Most such models neglect the connection between the torus and its surrounding.

A natural way to describe the accretion problem is to start from the inner disk, assumed in most models to be geometrically thin and optically thick, where the 
local viscosity
allows radial drift of material all the way to the innermost stable circular orbit (ISCO). A natural outer boundary of the accretion disk is its
 self gravity radius, $R_{SG}$: the location where local gravity exceeds the vertical component of the central BH gravity and the disk becomes unstable ($Q=1$).
This radius is given by:
\begin{equation}
 R_{SG} \simeq 1680 M_9^{-2/9} \alpha^{2/9} [L_{AGN}/L_{Edd}]^{4/9} [\eta/0.1]^{-4/9}  R_g \, ,
 \label{eq:r_sg}
\end{equation}
where $M_9$=\MBH/$10^9$\msun, $\alpha $ is the viscosity parameter which is of order 0.01-0.1, $\eta$ is the mass-to-radiation 
conversion efficiency, and $R_g$ is, again, the gravitational radius
\citep{Laor1989}.
For a stationary $M_9=1$ BH and a thin accretion disks radiating at  \Ledd$\sim0.1$, $R_{SG} \simeq 0.04$\,pc.

In the absence of additional local heating sources, $Q$ drops below unity beyond the self-gravity radius, and 
the disk is fragmented into clouds that move in the same general plane. The gap that is formed prevents the inflow of additional, more distant material
to the center.
According to some models \citep{Collin1999,Collin2008,Duschl2006}, 
a marginally stable situation with $Q\sim 1$ can be maintained over a large distance beyond $R_{SG}$ due to collisions between clouds, and SF and SN explosions that 
produce turbulence and local viscosity. This allows accretion from much larger radii. For very massive BHs, such a marginally unstable disk can extend to
$\sim 100 $ pc and beyond. A challenge of models of this type is to combine all these processes into a  coherent SF history which allows to follow 
the evolution of the nuclear star cluster.

Another comprehensive semi-analytic model by \citep{Vollmer2008}, considers gas infall from the galaxy which results in the
formation of a clumpy central disk. This leads to the formation of a torus through three phases of evolution:
1) An initial short and massive gas infall leading to the formation of a turbulent and stellar wind-driven $Q \sim 1$ disk. This is followed by energetic 
SF processes.
2) Once supernovae explode, the inter-cloud medium is removed, leaving a massive, geometrically thick, collisional disk with a decreasing,
but still high-mass accretion rate.  3) When the mass accretion rate has signiﬁcantly decreased, the torus becomes thin and 
transparent, similar to the circumnuclear disk in the center of the Milky Way \cite[e.g.][]{Alig2013}. 
Numerical hydrodynamical simulations by \cite{Schartmann2010} follow the final stages of the model focusing on BH feeding during phases 2 and 3.
The suggestion is that low velocity stellar winds, during the asymptotic giant branch (AGB) phase, starting some 50 Myr after the onset of SF, contain enough gas
to feed a small central BH for a period of about $10^8$ yr. The origin of such winds is a quasi-spherical star cluster with little angular momentum.
The winds collide, form a central thin disk, and the gas is accreted onto the BH via an unspecified viscosity mechanism.
 Such models are currently limited to small, $few \times 10^7$ \msun\ BHs.

A gaseous torus cannot retain a large height/radius ratio with pure rotation. Star clusters can retain such a shape
but not the gas between the stars since cloud collision will drive them to the plane on a few dynamical time scales.
Maintaining a thick gaseous structure requires high velocity turbulent motion, or outflow motion, and seems to work better in a clumpy medium where there 
are frequent  
cloud-cloud collisions \citep{Beckert2004}. Other proposed mechanisms to maintain a thick structure involve UV, optical and IR radiation pressure 
\citep[e.g.][]{Honig2007,Krolik2007,Czerny2011,Wada2012}, 
magnetic winds \cite[][]{Emmering1992,Konigl1994,Elitzur2006}, and SF in the inflowing gas.
UV-optical radiation by the central accretion disk, and locally  (inside the torus) emitted IR radiation, can be combined
into a plausible scheme that helps to solve the scale height problem. Here clumps are being pushed in the radial direction
by the central source radiation in a way that depends on \Ledd\ \citep{Honig2007,Wada2012}. 
The central source heats the dust in the clumps which in turn emits IR radiation. The local IR flux diffuses
outward being absorbed and re-emitted by neighboring clumps. This results in a radiation pressure
force that balances or overcomes the z-component of the central source gravity \citep{Krolik2007}.
In this picture, an active AGN is necessary to both produce the toroidal structure and maintain its thick structure.

The disk wind scenario proposed in \cite{Emmering1992} and discussed in numerous
other papers \citep{Konigl1994,Kartje1996,Bottorff2000,Elitzur2006,Elitzur2009}
 consists of a continuous large disk which is fed by cold material from the host galaxy.
 Close to the BH, it develops a magnetocentrifugal wind that supports rising clumps. 
In this scenario, the torus is merely a region in the wind that happens to
provide a toroidal-clumpy structure. Clumps inside
the sublimation radius are dust-free and can be viewed as BLR clouds. At larger distances, the clumps are dusty and can be viewed as part of
the torus. The exact column density of the clumps differ from one model to the next. It may depend on the magnetic field properties, 
the ratio of the external accretion rate to the mass outflow rate from the disk, and more.
The clumps that are most relevant to the observations are those capable of emitting strong lines, with a column density of 
at least $10^{22}$\,cm$^{-2}$. As the clumps rise from the disk, their column density drops dramatically and they disappear from view. 
The mass outflow rate in several wind models is somewhat smaller than the mass accretion rate through the disk. 

A  recent development is the understanding that dust can survive in the atmosphere of the inner accretion disk at
all distances where the disk effective temperature, $T_{eff}$, is below the sublimation temperature \citep{Czerny2011}.
Simple estimates show that at a distance of $R_{BLR}$ (the emissivity weighted radius of the BLR),
$T_{eff} \sim 1000$\,K, roughly the dust sublimation temperature. This raises the possibility that, starting from this distance, 
the {\it local radiation pressure} acting on the local grains,
is effective enough to raise dusty clumps from the disk surface. As the clumps
rise, they are no longer shielded from the central radiation field which is more intense than the local field,  
they are being pushed in a radial direction, and the dust evaporates.
This reduces the push in the vertical direction and the clumps fall back onto the surface of the disk. 
Further away, the grains in the lifted material
can survive the central radiation field and they form a clumpy thick structure.
This ``failed wind'' scenario is claimed to explain several BLR observations. 
Its main uncertainties are related to the nature of accretion disks
(thin, thick, etc), the variability of the central source radiation, and most 
importantly, the question of whether the local disk radiation pressure can support very large column density dusty clumps.

\begin{figure}
\includegraphics[angle=0,width=0.6\textwidth]{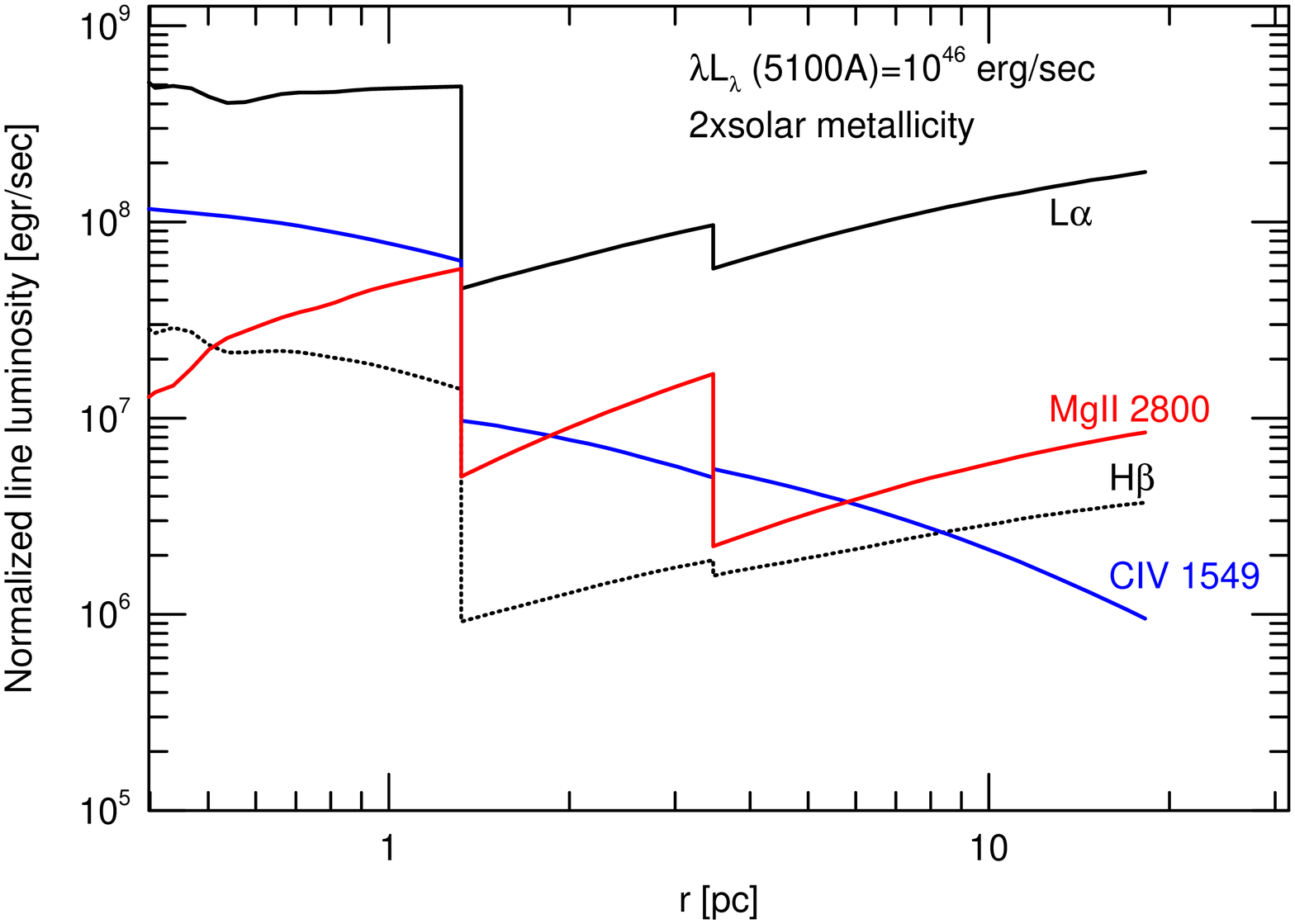}
\caption{
Relative line emissivities per unit covering factor for twice solar metallicity BLR gas
as a function of distance from the central BH. For the AGN luminosity considered here ($7 \times 10^{46}$ \ergs), the
graphite sublimation radius is at 1.32 pc and the silicate sublimation radius at 3.5 pc. These distances are clearly visible due to the large drop in the intensity of all lines resulting 
from the absorption of the ionizing radiation by the grains, and the selective depletion of most metals.
Note that some line emission is still present well within the dusty torus (for details of the model see \cite{Mor2012}).
}
\label{fig:mor_netzer_photoionization}
\end{figure}

Disk outflow models, magnetocentrifugal wind or radiation pressure driven wind, result in 
one continuous structure whose geometry is dictated by the global accretion process, and whose division into dusty and dust-free regions depends
entirely on the central radiation field. A major distinction between the dust-free clouds closer in (BLR clouds) and the dusty ones further away, 
is the efficiency of line formation. This can be 
calculated by photoionization models that provide the predicted emissivity of a specific broad emission line as a function of distance from the central
ionizing source. One example is shown in figure \ref{fig:mor_netzer_photoionization}. The parameters of the dust-free clouds in this model were chosen to
reproduces a typical broad line spectrum with the correct $R_{BLR}$. The diagram shows the dramatic decrease in line emissivity at 
the graphite and silicate sublimation radii due to the appearance of grains which absorb most of the ionizing flux. 
The details of the line emissivity depend on the run of ionization parameter but the global behaviour must be typical. Note
that line emissivities do not drop to zero and there is additional emission from the dusty part, inside the torus.

\begin{figure}
\includegraphics[angle=0,width=0.65\textwidth]{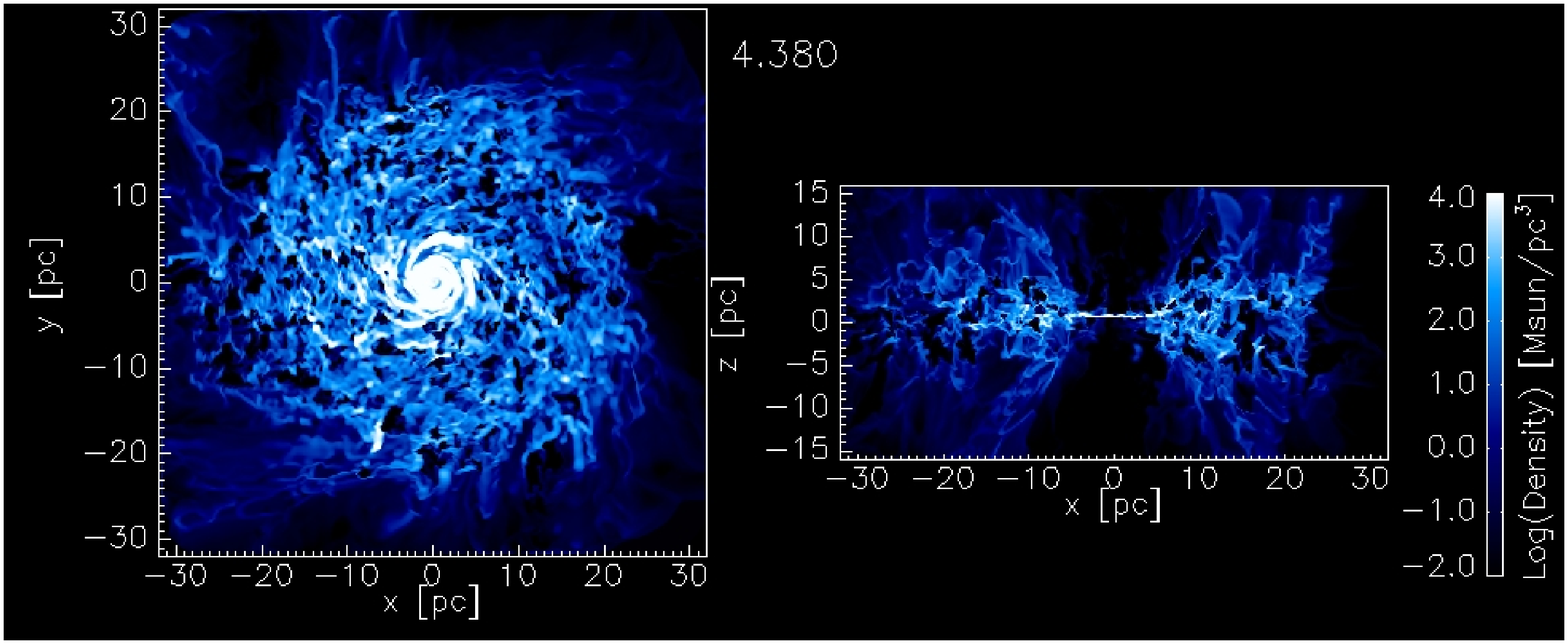}
\includegraphics[angle=0,width=0.3\textwidth]{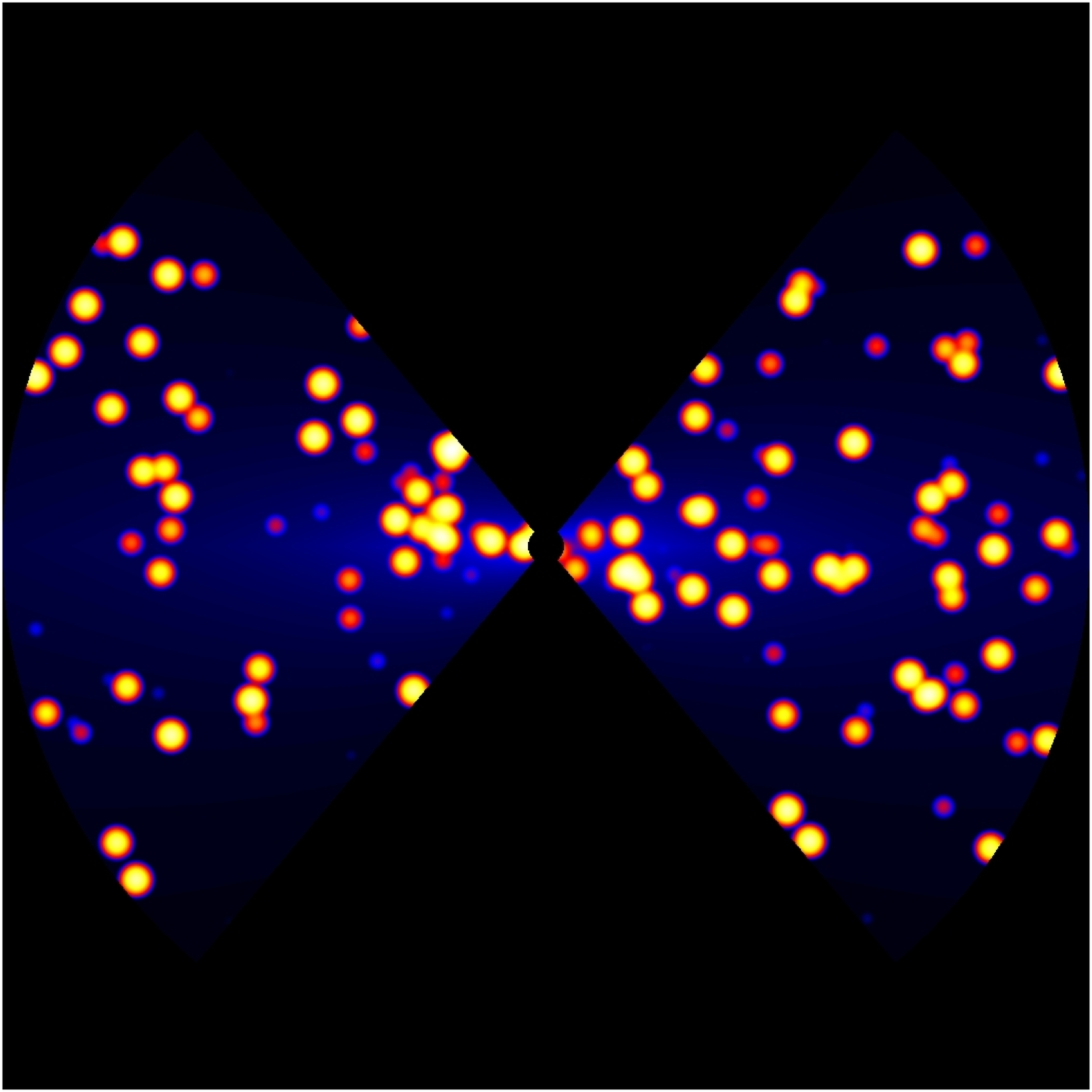}
\caption{{\it Left}: Face-on and edge-on density maps of a central toroidal structure resulting from gas infall towards a $10^7$\msun\ BH in a center
of a spiral galaxy as calculated by \cite{Wada2009}. Physical processes that are included
in the simulations are SF, SN explosions and stellar feedback (Courtesy of Keiichi Wada).
{\it Right:} Edge-on view of the 2-phase phenomenological model of \cite{Stalevski2012} showing 
high condensation clumps in yellow-orange colors and lower density inter-clump dust in
blue (Courtesy of Marko Stalevski)
}
\label{fig:wada_stalevski}
\end{figure}

Perhaps the most realistic models of today are numerical hydrodynamical simulations that include most but not all of the  processes discussed
above. 
The three-dimensional hydrodynamical simulations of \cite{Wada2009} and \cite{Wada2012} are perhaps the most detailed of their kind. They treat the ISM 
around a BH with $\sim 10^7$ \msun\ and assume SF and viscosity which increases during the SN phase. The 2009  work considers only stellar feedback.
It tracks  atomic and molecular hydrogen with a spatial resolution of 0.125 pc. In a quasi-equilibrium state,
 the gas forms an inhomogeneous disk with a typical diameter of about 30 pc. 
The velocity field of the disk is highly turbulent in the torus region and the
average SN rate is small but enough to energize the thick structure. The computed gas column 
densities are consistent with those derived by X-ray observations.
A more advanced version of the calculation \citep{Wada2012} includes AGN radiation pressure (AGN feedback). 
This drives a ``fountain'' and the
interaction between the non-steady outflow and inflow results in the formation of a geometrically thick turbulent torus.
Face on and edge on views of the density in one such model are shown in figure \ref{fig:wada_stalevski}

Present day numerical calculations have limited scope, in particular they treat low mass BHs and low accretion rate systems.
It is not at all clear that such scenarios are applicable to AGNs with BH mass of order $10^9$\msun\ and accretion rates of order 1\Msun/yr.
At high accretion rates, the local stellar population may be different, the SF events more violent, and the stellar population in the central
region different from the one observed in nearby systems..

It is important to note two other consequences of all torus models considered here.
The first is the connection of the torus typical dimensions with the gravitational potential of the galaxy which can be quantified by considering the
 ``sphere of influence'' of the central BH. 
This sphere determines the relative importance of the BH gravity and the bulge gravity, on material in the center. It is defined as
\begin{equation}
 R_{\rm BH,sph}=\frac{GM_{\rm BH}}{ \sigma_*^2} \simeq 10.7 \frac{M_{\rm BH}}{10^8 \msun} \left[\frac{\sigma_*}{ 200 \kms} \right ] ^{-2} \, {\rm pc} \, ,
 \label{eq:sphere}
\end{equation}
where $\sigma_*$ is the stellar velocity dispersion in the bulge.
For \Ledd=0.1, this corresponds to 
$ few \times 10^6 (\sigma_*/200{\rm km\,s^{-1}})^{-2} R_g$. 
For nearby AGNs, $ R_{\rm BH,sph} \sim 10$ pc, similar and even smaller than 
the predicted outer dimension of the torus. Thus, the gravitational influence of the stars in the bulge is comparable to that of the central BH and 
must be included in the calculations.  

The second consequence is related to broad line observations. All the models considered here require continuous flow of gas all the way from several 
pc to the ISCO.
This would mean blocking the view of the part of the BLR behind the disk in type-I AGNs that are observed close to face-on. 
The observational implications of this scenario have not been studied so far.
The masses in the torus and the central accretion disk are both of order 1\% of the BH mass, which, given an accretion rate that
produce \Ledd=0.1,  could supply enough gas for $few \times 10^6$ yr.
This raises the possibility that a long accretion episode of $\sim 10^8$ yr is broken into several shorter episodes during which 
the inner disk is disconnected from the outer flow.

An interesting variant of the general torus formation scenario, proposed by \cite{Wang2010} and \cite{Wang2011}, moves the SF activity to  
inside the dust sublimation radius. This model is not concerned with the issues of torus formation and stability.
It merely assumes the existence of a central torus which serves as a source of mass supply to the innermost region.
The model includes a thin, central accretion disk around the central BH, and fragmented clouds between $R_{SG}$ and the inner walls of the torus. 
SF activity and SN explosions take place inside the torus walls and the duration of one accretion episode 
is estimated to be $few \times 10^6$ yr.
This would result in $\sim 10^6$ stars that may eventually settle into a central star cluster. 

\subsection{Phenomenological torus models}

\subsubsection{Continuous and Clumpy Tori}
Phenomenological torus models is the name chosen here for ad hoc models that are constructed to enable a more accurate treatment of the radiative transfer, 
and  better modeling of the emitted SED and the images  of the torus at different wavelengths.
There are three generic types of phenomenological torus models: continuous (or smooth) gas distribution models,
\cite[e.g.][]{Pier1992,Granato1997,Efstathiou1995,Dullemond2005,Fritz2006}, 
clumpy torus models, \cite[e.g.][]{Dullemond2005,Nenkova2008a,Nenkova2008b,Honig2010}, 
and composite (a combination of clumpy and continuous) models \citep{Stalevski2012}. 
The \cite{Feltre2012} work is an illuminating comparison of the first two types and the work of \cite{Lira2013} provides  a comparison of two
clumpy torus models based on fitting the spectra of type-II AGNs.

All phenomenological models are axisymmetric and the torus inner walls are assumed to be at the dust sublimation
radius. Unfortunately, the definition of this radius can differ between models because of the different types of grains, 
 and the exact way used to estimate the bolometric
luminosity of the central sources.  Some of the models, e.g. \cite{Stalevski2012}, take into account the anisotropy of the central source radiation.
This adds an angular term to the torus equations ($f(\theta)$ in eqn.~\ref{eq:r_sub}) to describe the shrinking sublimation radius 
 at large polar angles. In this case, the torus equatorial plane reaches all the way to the outer boundary of the central accretion disk. 
The effect is more noticeable in the case of slim disks, where the angular dependence of 
the central source radiation is more noticeable \citep{Kawaguchi2010,Kawaguchi2011,Wang2014}.

The important parameters of the continuous phenomenological models are the inner and outer radii of the torus, the density profile and the opening angle.
Additional parameters of clumpy tori models are the column density of individual clumps, the radial distribution of the clumps (this replaces the run of
gas density in the smooth model), the filling factor and density distribution of the clumps. Present day phenomenological 
models do not yet include different types of  
grains in different parts of the torus although this has been considered in simpler (from the point of view
of radiative transfer) models like \cite{Schartmann2008}.

Radiative transfer is relatively easy to handle in continuous dusty tori but this is not the case in clumpy tori.
Different studies adopt different approaches to solving this problem. The analytical 1D calculations of \cite{Nenkova2008a} are limited in 
their ability to treat complicated geometries but can cover a large parameter space with different
clump sizes, torus opening angle and clump distribution. The 3D calculations of \cite{Honig2010} and \cite{Stalevski2012} are example of models using 
a Monte Carlo (MC) transfer technique, combined with ray-tracing, that allows more accurate solutions and, in principle, can deal with different grain size
 and composition.  However, the calculations are computationally intensive and the parameter space explored more limited.

Continuous dust distribution models are characterized
by a monotonically decreasing  dust temperature as
a function of distance from the BH due to the $F \propto r^{-2}$ term and the attenuation of the incident 
radiation by the dust. In such models, every radius corresponds to 
a certain temperature.
One prediction of continuous models is strong silicate 9.7\mic\ absorption in cases where the total optical depth at
this wavelength exceeds unity. This is simply the result of the absorption of the flux emitted at this wavelength by 
the inner torus walls by further away material. This is predicted to be most noticeable in type-II sources that are seen
at large inclination angles \cite[e.g.][]{Nenkova2008a}.

There are two fundamental reasons for the different SEDs of clumpy torus models. First, the illuminated and back sides of individual clumps
radiate at very different temperatures. This allows a certain location in the torus to have a large range of temperatures rather than the
single temperature typical of continuous models. Second, and more important, gaps between the clumps allow a free, un-attenuated 
line of sight from the center to deep inside the torus. A clump at several $r_{sub}$ can emit at the local, distant dependent dust temperature
spreading the high temperature regions over larger parts of the torus.  
There are two important consequences: the strength of the 9.7\mic\ absorption feature is much below its predicted strength in continuous
distribution cases, and the anisotropy of the emitted radiation is smaller.
An obvious limitation is the assumption of empty space between the clouds in a system where cloud collision and evaporation are unavoidable.

\begin{figure}[th!]
\includegraphics[angle=0,width=0.75\textwidth]{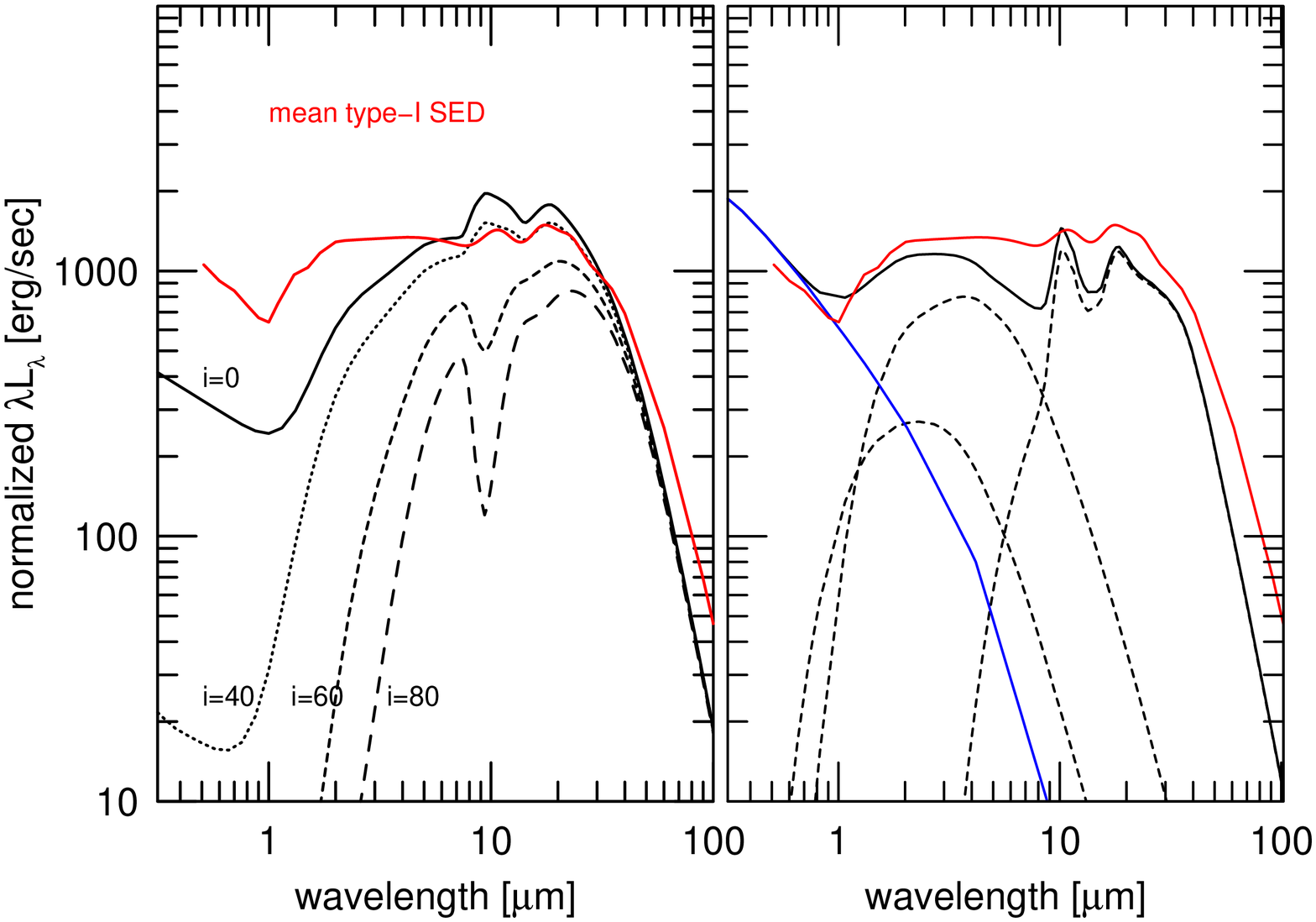}
\caption{{\it Left:} A comparison of the \cite{Stalevski2012} models (black lines) with the observed \cite{Mor2012} composite type-I SED (red line)
that contains dust emission from the NLR (see \S\ref{sec:torus_observations}). The models are
marked with the viewing angle to the torus.  {\it Right:} The SED of a three cloud obscurer. The column densities of the clouds 
are $10^{23}$ cm$^{-2}$ and the gas-to-dust ratio galactic. Two of the clouds contain pure graphite grains and are located at
the graphite sublimation radius and three times further away. The third cloud contains ISM-type dust and is located at a distance which is
100 times further than the first cloud. The covering factors are 0.1, 0.3 and 0.3, respectively. 
The solid black line is the combination of the three individual SEDs that are
marked in dashed lines. The blue line is the SED of the central disk.
}
\label{fig:netzer_stalevski}
\end{figure}

The third type of composite models is, perhaps, the most realistic. In this case, the volume between clumps is filled with diluted dusty gas which absorbs 
part of the incident optical-UV radiation and part of the locally emitted NIR and MIR radiation.
The main features of the pure clumpy case are preserved but additional attenuation by the inter-cloud dust and gas must be taken into account.
Fig.~\ref{fig:wada_stalevski} shows a schematic structure of the composite torus model of of \cite{Stalevski2012} and figure \ref{fig:netzer_stalevski}
exhibits several SEDs computed with this model. The SEDs are compared to the composite type-I SED of \cite{Mor2012}.

To illustrate the difficulty in deriving specific torus properties from comparing models with observations, I show in
Figure \ref{fig:netzer_stalevski} a very basic model that was constructed by combining the emitted fluxes of three (!) dusty clouds with the same column
density, $10^{23}$ cm$^{-2}$. The clouds are located at different distances from an accreting BH. Two of them contain pure graphite grains
and the third graphite and silicate grains similar to what is found in the ISM
(see details in the figure caption). This simplest possible theoretical SED is already in 
reasonable agreement with the observations suggesting that doubling or tripling the number of clouds is all that is required to get a very good fit.

Finally, we should note two other types of phenomenological models, the tilted disk model of
\cite{Lawrence2010} and the combined torus plus dusty polar wind model of \cite{Honig2013}. 
The first is discussed in \S\ref{sec:additional_obscurers} and the second in \S\ref{sec:torus_observations}

\subsubsection{Anisotropy and Covering Factor}

Each of the models discussed above has its own effective covering factor.
For smooth torus models, this is simply a matter of geometry and the wavelength dependent optical depths. For clumpy models, it depends on the
number of clouds and their 3D distribution. In the straw person model of \cite{Antonucci1993}, the covering factor is assumed to be the same for
all sources of a given luminosity. The implication to large samples is translated to a simple expression: \Cf=N(type-II)/N(type-I).
The realization of a significant spread in \Cf\ at a given luminosity,
suggests that the covering factor is a second unification parameter \citep{Elitzur2012}. In this case, the relative number of type-I and type-II 
sources still defines the mean \Cf\
but  a source  is more likely to be classified as type-II if its covering factor
is larger than the mean and as type-I if it is smaller than the mean.
 Thus the  high covering factor tail of the \Cf\ distribution contains more type-II AGNs, and the mean  
covering factor of type-Is is smaller than the mean of type-IIs if objects are selected by their optical properties.
More quantitatively, if we define the covering factor probability distribution of sources with luminosity $L$ as
$P$(\Cf,$L$), and the probability for a type-I source from this population to have a covering factor \Cf,  $P_I$(\Cf,$L$),
we can write:
\begin{equation}
  P_I(f_C,L) = (1-f_C) P(f_C,L) \,\,\,\, ; \,\,\,\,  P_{II}(f_C,L) =f_C  P(f_C,L) \, .
\label{eq:p_cf}
\end{equation}
Because of the finite probability of seeing the central source at {\it any inclination} 
through a clumpy medium (assuming no dusty inter-clump material),
some sources classified spectroscopically as type-I AGNs, have high inclination angles. Similarly, an obscuration by a single large cloud 
can result in the appearance of a type-II spectrum. This interesting possibility can be tested by observations. For
example, the angle of polarization of hidden broad emission lines in face-on tori, is predicted to be very different from edge-on tori.
The presence of a lower density, dusty intercloud medium \citep{Stalevski2012} changes this simple estimation but the overall dependence on the angular 
distribution of the clumps is still an important feature of all clumpy torus models.

Anisotropy is a second characteristic of all torus models. This is
most noticeable in continuous gas distribution tori where the smallest-distance dust, which is directly heated by the central source, 
can only be seen from small inclination angles. In clumpy tori, there are always gaps in the distribution and hot clumps can be seen at larger
inclination angles. This reduces the anisotropy. 
Inspection of several of the \cite{Stalevski2012} and \cite{Nenkova2008b} models designed to fit observed SEDs,
suggests that for wavelengths that are optically thick
to the torus emitted radiation, the anisotropy is roughly proportional to $\cos i$, where $i$ is the inclination to the LOS. This is similar
to the expected anisotropy of the central accretion disk.
At longer wavelengths, where the dust optical depth is smaller, the emission is more isotropic and at FIR wavelengths
the torus emits more like a sphere.

A general, somewhat simplistic approach which does not take into account wavelength dependences is, 
nevertheless, useful for improving the estimated geometrical covering factor
for type-I AGNs. This approach requires an isotropy parameter $b$ that varies from 1 (complete isotropy) to 0 (maximum anisotropy),
and a general dust anisotropy function, $a(b,f_C)$.
Assume a case where the (unknown) torus covering factor is \Cf\ and the total AGN bolometric luminosity, \LAGN, is completely isotropic
 and seen, unattenuated, through the opening in the torus, i.e. (1-\Cf)\LAGN\ escapes the system.
The total dust
emission is $f_C$\LAGN\ and the emission through the opening is 
\Ldust=($1-b f_C)f_C$\LAGN. 
Recall that flux measurements refer to luminosity per unit solid angle (i.e. divide the AGN and dust emission through the opening by (1-\Cf)), we can define 
an observed ratio,  $R$=\Ldust/\LAGN\ and relate it to the covering factor by requiring
$R=a(b,f_C) f_C$, where the anisotropy function is:
\begin{equation}
a=\frac{1-b f_C}{1-f_C} \, . 
\label{eq:covering_anisotropy}
\end{equation}
This gives $a(b,f_C)=1$ for complete isotropy ($b=1$) and $a(b,f_C)= 1/(1-f_C)$ for maximum anisotropy ($b=0$), as required. 
For example, in a case where \Cf=0.7, complete isotropy of dust emission gives $R=f_C$, maximum anisotropy gives $R= 3.33 f_C$,
and an intermediate case
with $b=0.5$ gives $R= 2.17 f_C$.

Finally, in optically thick tori, the covering factor may also be related to the narrow line intensities since the smaller the
covering factor is, the larger is the opening and more NLR gas is exposed to the central source radiation.
As shown in \S~\ref{sec:ionization_cones} below,
this is a somewhat simplistic approach since the NLR gas
does not necessarily fill the entire cone  and since absorption by dusty
inter-clump material close to the edges of the torus cannot be neglected.
Moreover, the intensity of the narrow emission lines relative to the bolometric luminosity, \LAGN,  increase with decreasing source luminosity 
in a way which seems to be completely unrelated
to the torus covering factor.

\section{TORUS OBSERVATIONS}
\label{sec:torus_observations}

The large number of successful IR space missions of the last 20 years, the on going successful X-ray missions,
and the significant improvements in ground-based interferometry, revolutionized the study of AGN tori.
They provide huge data sets that can be used to obtain reliable IR SEDs, allow multi-wavelength studies of many thousands of AGNs,
and enabled to probe regions that are 1 pc in size near
the BH.
This section describes these developments and their application to individual AGNs. The related redshift and luminosity dependences 
are discussed in \S\ref{sec:unification_or_evolution}.

\subsection{The Inner Torus Boundary: Dust Reverberation Mapping}
The variable luminosity of the central source results in time-dependent heating and ionization of the surrounding gas and dust that can be used to obtain
valuable information about the geometry and physical conditions in these regions. 
Reverberation mapping (RM) has been used, for years, to map the distribution of the line emitting gas, mostly \hb, in the BLR 
\cite[``line RM'', see][and references therein]{Kaspi2000,Bentz2013}. Other experiments were designed to measure the time-dependent dust emission in the
NIR, mainly in the  K-band, in response to the variable continuum \cite[``dust RM'', see e.g.][]{Glass1992,Suganuma2006,Kishimoto2007,Koshida2014,Pozo2014}.
The K-band wavelength, $\sim 2.2$\mic, is close to the peak emissivity of the hottest dust and the experiments provide information about
the inner torus boundary (\S\ref{sec:central_obscurer}).

\begin{figure}[ht!]
\includegraphics[angle=270,width=0.6\textwidth]{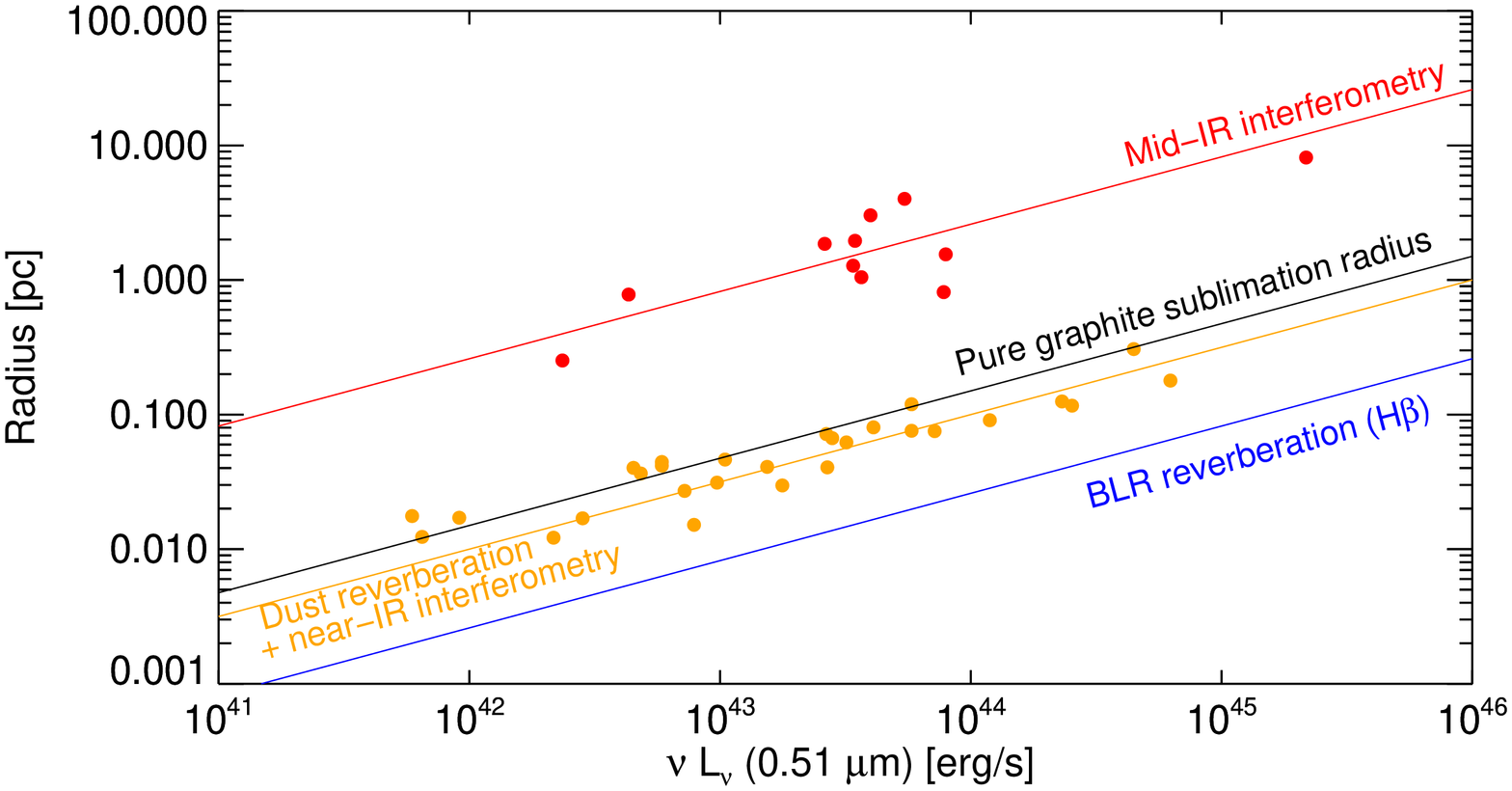}
\caption{
Various measures of the size of the central torus:
MIDI-based measurements of $R_{1/2}$  (red points and a red line),
dust RM measurements based on V and K-band light curves and K-band interferometry (orange points and an orange line), graphite sublimation radius 
from eqn.~\ref{eq:r_sub} (black line),
and the \hb\ RM size (blue line).
 The Typical 8--13\mic\ size is $\sim 30 $ times larger than the
typical dust RM size.
(adopted from \cite{Burtscher2013}, courtesy of Leonard  Burtscher).
}
\label{fig:leonard_MIDI}
\end{figure}

The main results of dust RM experiments, now carried out successfully in about 20 low-to-medium luminosity AGNs, is a tight correlation of the form:
$R_{\rm 2.2 \mu m}\simeq 0.4 (L_{46})^{1/2}$ pc \cite[][]{Koshida2014}, 
where $L_{46}=$\LAGN/$10^{46}$ erg~s$^{-1}$  
and I converted the measured V-band luminosity to \LAGN\ assuming   
\LAGN$\simeq 8 \lambda L_{\lambda}(5500)$\AA.
Thus, to a very good approximation, $R_{{\rm 2.2 \mu m}}=\langle R_{\rm sub,C} \rangle$  (eqn.~\ref{eq:r_sub}).
Note, however, that different values for the dust sublimation radius have been used in the literature \cite[e.g.][]{Koshida2014} which lead to 
suggestions that the measured V-K time-lag is 2--3 times smaller than expected from sublimation temperature considerations.
The comparison of the \hb\ and K-band RM results is shown in Figure \ref{fig:leonard_MIDI}.
The measured K-band time lag is a factor 3--4 larger than the \hb\ time lag, a result which is expected given what we know about gas and emissivity distributions
in the BLR \citep{Netzer2013}.

RM experiments are not without difficulties. The dust transfer function (the dust emission in response to a delta-function central source light-curve) 
must be complicated, and differs substantially from simple thin or thick shell. 
It depends on the torus and disk inclination, the dimming of the disk radiation towards its equatorial plane, the occultation of some parts of the torus, 
and more \citep{Kawaguchi2010,Kawaguchi2011,Pozo2014}.
Nevertheless, the agreement between the dust RM results, and the predicted graphite sublimation radius, suggest that the inner torus wall is 
determined by the maximum  grain temperature.

Two comments about the relative sizes of the BLR and the inner torus walls are in order.
The \hb\ radius measured by RM experiments should not be mistaken for a
single location BLR. In fact, the more successful BLR models require the ionized dust-free gas to extend all the way from the inner 
disk to the torus \cite[e.g.][]{Mor2012}.
In addition, the quoted distance ratio of 3-4 is only an average and actual ratios range from 1.5 to about 10.
 This reflects the uncertainties on individual measurements and the likely large scatter in RM-based distances.

\subsection{Torus Properties: SED fitting}

The various torus models of \S\ref{sec:torus_models} make specific predictions about the emitted SED which can
be compared with NIR-MIR observations. This has been an active area of research 
because of several successful space missions, mostly the Spitzer space telescope (\spitzer) and the wide-field infrared survey explorer (\WISE).
The satellite data are supplemented with  ground based, J, H, K, L and N-band observations with inferior  wavelength coverage
but much superior spatial resolution.
The spectroscopic characteristics of the torus that allow to distinguish type-I from type-II SEDs are 
the 9.7 and 18\mic\ silicate features and the 1--8\mic\ continuum slope. 
The comparison between observations and theory discussed below is biased towards phenomenological models since these are the ones that were
used in almost all such studies.

The spatial resolution of the IR observations is of great importance because of galactic scale emission that dilutes the characteristic signature
of the torus. Stellar emission can contribute to the observed signal in the NIR and dust emission from SF regions can dilute the 
torus signature at 10--30 \mic. 
In particular, strong polycyclic aromatic hydrocarbon (PAH) features, mostly at 6.2, 7.7 and 11.3\mic, can contribute significantly to the MIR 
emission and dilute broad band photometric observations. In addition, dust emission from the NLR can contribute significantly to the 10--30\mic\ continuum
\citep{Schweitzer2008,Mor2009,Mor2012}. Because of this, small aperture spectroscopic observations are preferred. Unfortunately, the \spitzer\
spectroscopic apertures are several arcsec in size 
and include significant contribution from the host galaxy in low luminosity AGNs (see a large collection of those in \cite{Buchanan2006}).
A detailed comparison of large and  small aperture spectroscopic observations is given in \cite{AlonsoHerrero2011}
and \cite{Lira2013}.

The works of \cite{RamosAlmeida2011} and 
 \cite{RamosAlmeida2014} are good examples of high spatial resolution ($\sim 35$ pc), very detailed observations of a very small number of objects.
The results are similar to other works by this group and claim to be ``clean'' since 
they avoid galaxies with known  dust lanes that affect the SED shape and  the depth of the 9.7\mic\ feature.
The 1.5--19 \mic\ SEDs obtained in those studies indicate a considerably different slope for the two types of AGNs:
For type-Is $\lambda L_{\lambda} \propto \lambda^{0.7}$, and for type-IIs  $\lambda L_{\lambda} \propto \lambda^{1.6}$, with a rather large scatter
in both slopes.  
However, the \cite{Lira2013} work on a much larger sample of type-II sources, disagrees with some of the results. It 
shows a very large range of slopes, including some that are similar
and even flatter from the type-I slopes found by \cite{RamosAlmeida2014}. It seems that there is no canonical NIR-MIR slope for
low luminosity type-II AGNs.

Most attempts to derive torus parameters, such as the number of clumps along a certain direction and clump optical depths, give inconclusive results.
\cite{Lira2013} made a detailed comparison between the rather different clumpy torus models of \cite{Nenkova2008b} and \cite{Stalevski2012} 
and were able to obtain statistically satisfactory fits for both. 
One of the parameters of the fit is the inclination angle but, unfortunately, there is no correlation between the angles found by
applying the two models.
\cite{Lira2013} also found that SED fitting of about half of their type-II sources requires an additional very hot dust component, with dust temperature 
of $\sim 2000$\,K, to 
explain the observed NIR flux. Such hot dust is not included in any of  the clumpy models used in this work.
It seems that the NIR-MIR spectra of AGNs of all types do not contain enough spectral features to distinguish one object from
the next, and one clumpy torus model from another. In other words, torus models 
contain enough free parameters to fit almost any observed SED.

\begin{figure}[ht!]
\includegraphics[angle=0,width=0.6\textwidth]{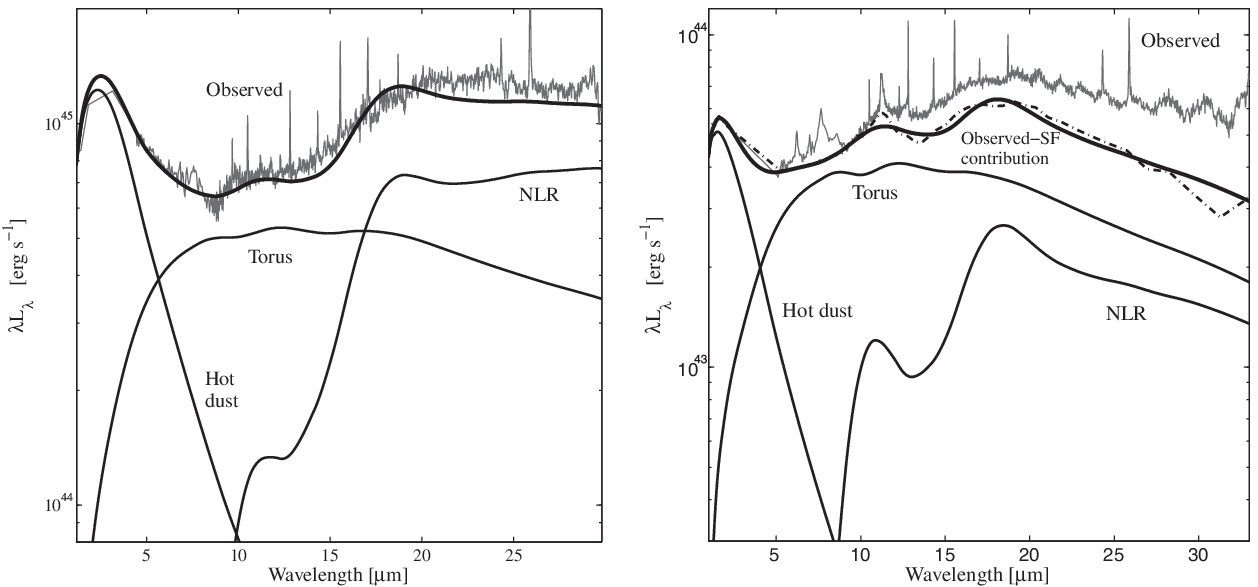}
\includegraphics[angle=0,width=0.4\textwidth]{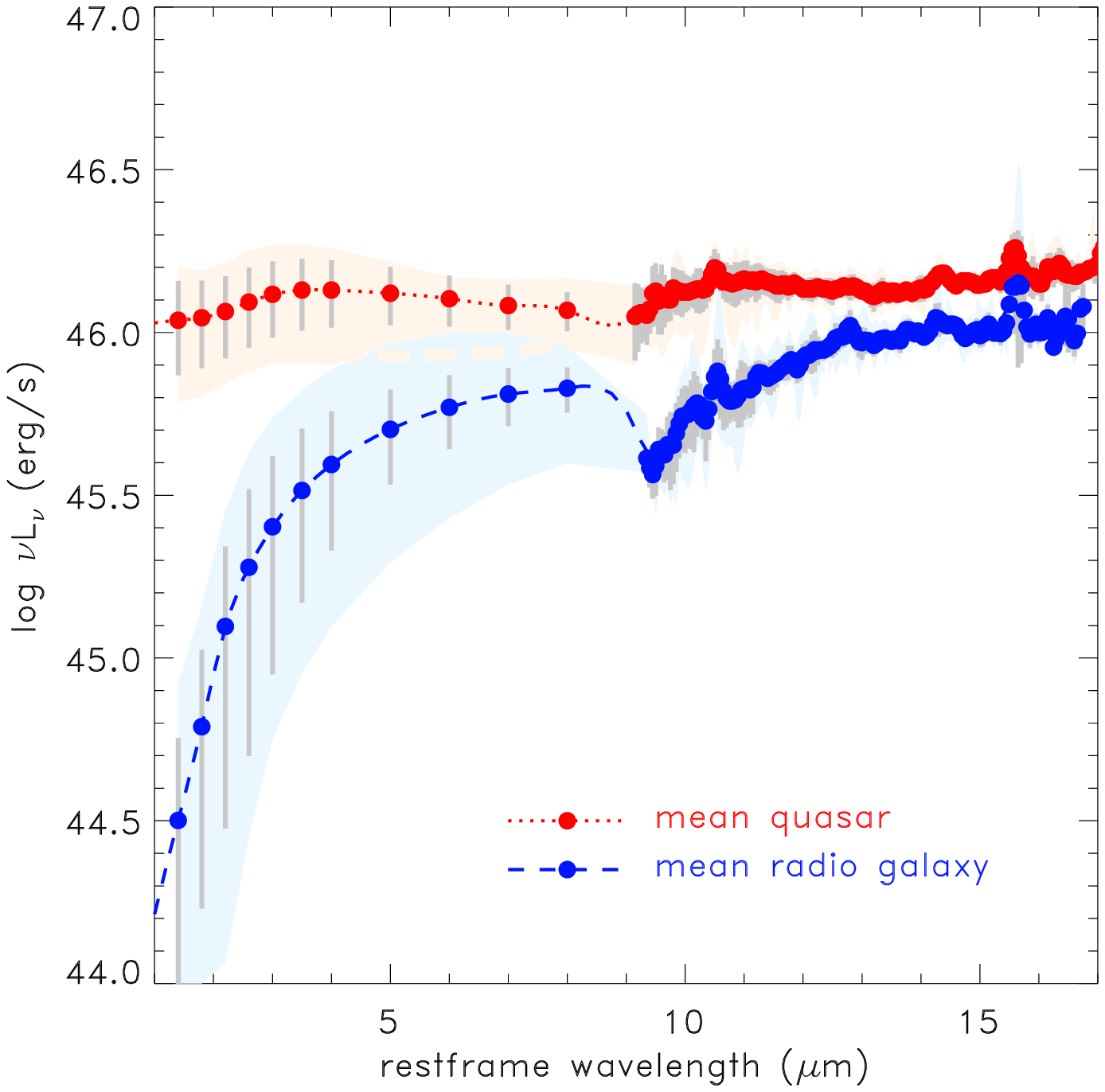}
\caption{
{\it Left:} Observed (gray) and SF subtracted (dot-dashed line) SEDs of two type-I AGNs from \cite{Mor2012}.
The diagrams show the three components used in the fit:  hot pure graphite dust, a clumpy
torus model with ISM-type dust, and dust emission from the NLR. The thick black curve is the sum of the three.
{\it Right:} Composite type-I (quasars) and type-II (radio galaxies) SEDs of $z\sim 1.2$ steep spectrum radio-loud AGNs from \cite{Honig2011a}
illustrating the large spectral differences between the two groups.
}
\label{fig:mor2012}
\end{figure}

There are several examples of SED fitting of luminous, type-I AGNs \cite[][]{Schweitzer2008,Mor2009,Roseboom2013,Assef2013,Lusso2013}.
\cite{Schweitzer2008} demonstrated that successful SED fitting of large aperture observations requires, on top of the standard torus model, 
two additional components:
very hot, T$>1500$ K emission from the torus and warm dust emission from the NLR. This approach was later used by \cite{Mor2012}
who combined the \cite{Nenkova2008b} clumpy torus model with two additional components, one containing pure graphite grains and the other a dusty
NLR. The hot dust component is not an integral part of the \cite{Nenkova2008b} models and its treatment is rather simple.
After subtracting the host SF contribution, they fitted \spitzer\ spectra combined with NIR photometry and obtained satisfactory fits, 
like the ones shown in figure \ref{fig:mor2012}, for more than a hundred medium and high luminosity AGNs. 
The composite SED of this sample (red line in  figure \ref{fig:netzer_stalevski})
indicates a very flat, slope zero (in $\lambda L_{\lambda}$) 2--20\mic\ continuum and a turn down at longer wavelengths. The exact long wavelength dependence is very
sensitive to the assumed SF contribution. Indeed, a parallel effort by \cite{Mullaney2011}
using a smaller sample of AGNs of both types, suggested a decline that starts at a longer wavelengths of about 30\mic. 
As shown earlier (figure \ref{fig:netzer_stalevski}), a very simple combination of very few large clumps can fit such observations almost as well
as the sophisticated clumpy models.

The \cite{Mor2012} SED fittings was used to put constraints on
the various parameters of the clumpy torus model of \cite{Nenkova2008b}.
The results show a well defined inner radius, an ill defined outer radius, and a broad distributions of cloud
properties.
Many of these properties are similar to those found by \cite{Lira2013} when fitting their type-II objects.
The fittings also allow to compare the three major contributions to the total dust emission.
For the luminosity median of the population, L(hot torus dust):L(warm torus dust):L(NLR dust)$\approx$40:40:20. The relative NIR and MIR luminosities
 in the \cite{Roseboom2013} sample, that does not consider the  NLR contribution, are similar and somewhat luminosity dependent. The fits also
provide an estimate of the luminosity dependent size of the NLR, 
\begin{equation}
R_{\rm NLR} \simeq 460 \left[ \frac {L_{\rm AGN}}{ {\rm 10^{46}\,erg\,s^{-1}}} \right ]^{0.67} \,\, {\rm pc}\, .
\label{eq:mor_nlr}
\end{equation}
This dimension is more that two orders of magnitude larger than the silicate sublimation radius and 
is in agreement with the inner part of the NLR derived from direct imaging and and photoionization models.
It corresponds to a mean grain temperature of $\sim 160$ K with peak emission at around 18\mic.

An additional detailed study of the NIR-MIR SED of $z \sim 1.2$ luminous radio-loud AGNs, by \cite{Honig2011a}, covers the rest-frame wavelength 
of 1--17\mic. The sample includes 3CRR sources that were originally
described in \cite{Haas2008} and \cite{Leipski2010b} and are very similar in their IR-optical-UV properties to radio-quiet AGNs.
 Despite its small size (20 objects), the sample is most suitable for the comparison of type-I and type-II 
sources since it is chosen on the basis of the extended radio emission which is expected to be radiated isotropically.
While the mean radio luminosity of the two groups is almost identical, the NIR-MIR spectrum is very different
(see figure \ref{fig:mor2012}). The type-I SEDs (referred to as ``quasars'' in the diagram)
are flat in $\lambda L_{\lambda} $ and very similar to the \cite{Mor2012} composite.
The type-II SEDs (referred to as radio galaxies) are steeply rising from 1 to 7 \mic\ and are similar to some of the type-II SEDs discussed earlier.

In summary, the typical NIR-MIR SEDs of type-I and type-II AGNs are rather different. Most, but not all type-Is show 
significant hot dust emission. Type-IIs show a much larger diversity with some objects similar to 
type-Is and others showing a steep decline from long to short wavelengths. Clumpy torus models provide adequate fits to many observed SEDs 
but seem to be degenerate
with respect to their basic parameters. Because of this, models with different geometry, dust distribution and even inclination, 
can produce equally good fits to the same data. 
The inner region of the torus which contains pure graphite dust is clearly required by the observations but is  missing from todays models.  
Given all this, it is not yet clear that current phenomenological models resemble, well, real AGN tori.

\subsection{Anisotropy and Covering Factor}
 
As explained in \S2.3.2, dust emission from optically thick tori is characterized by a wavelength dependent anisotropy. 
Several of the torus models described in \S\ref{sec:torus_models} are optically thick through most of the 1--20\mic\ wavelength range
and  SED differences between type-I and type-II sources can be used to derive the anisotropy.   
 According to \cite{Honig2011a}, this is the cause of the different SEDs shown  in the right panel of figure \ref{fig:mor2012}.
 In this particular case, if the type-II inclination angle is 75 degrees, and the type-Is 39 degrees, 
the column density needed to produce this anisotropy corresponds to 
 $A_V\sim 150$ mag. Even at such a large column, the optical depth is small and the emission is isotropic at $\lambda \geq 30 $\mic.

A direct way to estimate the anisotropy is to measure the dust optical depth through sensitive search for broad Paschen and Brackett lines in 
the spectrum of type-II sources. The total number of objects studied in this way is small \cite[see][and references therein]{Lutz2002} and the 
typical detection
rate is only about 30\%. The most useful results were obtained for type-II AGNs showing broad Balmer lines in polarized light
indicating a hidden BLR. The hand-full of broad Br$\alpha$ lines
reported by \cite{Lutz2002} indicate that $\tau (4.05 \mu m) \sim 3$ mag., corresponding to column densities of order $10^{23}$ cm$^{-2}$ that are
consistent with the columns derived from X-ray observations of the same sources. For several 
objects in this sample, with polarized broad emission lines but no direct broad line detection at 4.05\mic, the obscuration must be larger.
As shown below in \S\ref{sec:additional_obscurers}, such a good agreement is not typical of many
AGNs.

The large NIR optical depths deduced from the Br$\alpha$ observations should 
not be confused with the observations of many type-Ii AGNs showing weak broad \ha\ lines and basically no broad \hb\ indicating
a few magnitudes of visual extinction.
Such objects are common with numbers that are estimated to be 10--20\% of all low redshift AGNs 
\cite[e.g.][see \S\ref{sec:the_agn_family}]{Lawrence1982,Antonucci1993,Elvis2012}. They are
thought to be reddened by galactic scale dust although in some cases, the source is probably seen through a small column density 
part of the torus.
Reddening of high luminosity high redshift AGNs is also a common phenomenon \cite[e.g.][]{Glikman2004,Lacy2007}. Such objects are found in IR and
radio surveys and their weak, occasionally undetected optical emission suggests a large range of dust extinction, from that typical of many local type-Ii sources,
to much higher. Their fraction in the general population must be high but is not accurately known

As already alluded to, anisotropy  influences covering factors estimates based on the measured \Ldust/\LAGN\
(eqn.~\ref{eq:covering_anisotropy}).
The estimation of the mean and the distribution of \Cf\
requires an additional correction that takes into account the fact that a given \LAGN\ is associated with a range of covering factors.
In particular, a large \Cf\ increases the probability of classifying a source as 
type-II and the opposite is true for type-Is (eqn.~\ref{eq:p_cf}). 

Covering factors for a large number of type-I AGNs  have been obtained by using the \Ldust/\LAGN\ method
\cite[][]{Maiolino2007,Treister2008,Mor2011,Mor2012,Roseboom2013,Lusso2013}. The quality of the SED fits in such studies
ranges from poor to very good,
where spectroscopy-based fits provide more reliable measurements. Not all studies
include the anisotropy correction term which results in overestimation of \Cf.
All studies show a broad distribution of \Cf\ for a given luminosity, with a type-I population mean of 0.3-0.6. 
In general, there is clear indication for a decrease of the mean covering factor with increasing source luminosity.
Some of the studies attempted to separate hot (NIR) from warm (MIR) dust
\citep{Mor2011,Mor2012,Roseboom2013}. They show that the hot-dust covering factor varies over a larger
range with some sources missing this component all together (``hot-dust poor AGNs'', see \cite{Mor2011}).
The fraction of such sources among luminous AGNs is about 15\%.
In general \Cf(hot-dust)$\approx$\Cf(warm dust) but the separation of the two is model dependent.
A big disadvantage of this method as applied to IR-selected samples is the bias against low covering factor sources.

The torus covering factor can be obtained directly from modeling the observed X-ray continuum (see \S\ref{sec:the_agn_family}).
This has been demonstrated by \cite{Ueda2007}  and by \cite{Brightman2012} who
made a careful comparison of the scattered and transmitted X-ray flux of a large number of type-II AGNs. They
claimed to be able to measure scattered radiation which is as low as 1--2\% and thus identify  
``buried'' active BHs that are completely surrounded by geometrically thick tori with \Cf$\simeq 1$.
The work suggests that some 20\% of high redshift sources in several
deep X-ray fields are completely buried AGNs.

An additional way to estimate \Cf\ is based on the observed intensity of the narrow Fe~\Ka\ line
(\S\ref{sec:central_obscurer}). Studies of large X-ray samples show
that EW(\Ka) decreases with increasing X-ray luminosity, roughly like \LAGN$^{-0.15}$, but not with redshift
 (the ``X-ray Baldwin effect'' \cite[see e.g.][]{Chaudhary2010,Ricci2013}). As explained, EW(\Ka) depends on the combination of
the torus column density,
covering factor and metallicity \citep{Ikeda2009,Yaqoob2010,Brightman2011a,Ricci2013}.
The advantage of this method is that the selection of
sources by their X-ray flux is independent of the measured quantity, EW(\Ka). 
The possible complications are extra emission from \ka\ produced in broad line clouds and the fact that EW(\Ka) depends also on
the column density.
Moreover, the measured LOS X-ray column, and the mean column density of the torus, are not 
necessarily the same \cite[see e.g.][]{Brightman2012}.

Finally, a complementary approach to find a population mean covering factor is to measure the LOS obscuring column and associate 
small columns with type-I sources and large columns with type-II sources. 
The mean \Cf\ is then obtained from the ratio N(type-II)/N(type-I), where N stands for the number of objects.
This method is discussed in \S\ref{sec:unification_or_evolution}.

It is interesting to examine the covering factor of the most luminous AGNs.
An example is the work of \cite{Weedman2012} who compared the  
\WISE\ (rest-frame 7.8\mic) and SDSS (rest-frame wavelength of 1350\AA) luminosities of the most MIR luminous type-I AGNs
at redshift 1.5--5.
The study assumed \Ldust/L(7.8\mic)$\simeq 3$ and found 
extremely high values of \Ldust, $10^{46.9-47.3}$\,\ergs,
and \Cf$>1$ for many sources.
The problem with this and several other studies is the neglect of known selection effects and necessary correction factors.
like the anisotropy correction term of eqn. \ref{eq:covering_anisotropy} and the possible
anisotropy of the accretion disk radiation. 
Moreover, these objects are probably unique among type-I AGNs in terms of
their covering factor because they were selected to be the most IR luminous.
Indeed the agreement between the luminosity distributions at 7.8\mic\ and at 1350\AA, is very poor.
Finally, a large fraction of the sources may be reddened, because of the large dust covering factor, 
a possibility that was acknowledged by the authors.
Taken together, it seems that the mean torus covering factor for the most luminous type-I AGNs is $\sim 0.5$.

An important way to evaluate {\it the combined effect} of anisotropy and covering factor is to compare the X-ray and MIR luminosities, 
L(MIR)/L(2--10 keV).
The comparison made at 6 and $\sim12$\mic\ shows an almost identical mean ratio for
the two AGN types, over four orders of magnitude in luminosity, with a typical scatter of about 0.3 dex 
\citep{Lutz2004,Levenson2009,Gandhi2009,Asmus2011,Honig2010,Brightman2011b,Gonzales2013,Merloni2014}. 
This is a somewhat surprising result given the covering factor and anisotropy discussed earlier.
Naive expectations suggest that the larger covering factor of type-II sources will increase 
L(MIR)/L(2--10 keV) relative to type-Is. On the other hand, anisotropy of the MIR emission
would tend to decrease this ratio in type-II AGNs that are typically seen at higher inclination angles.
Emission by NLR dust will affect large aperture observations and will contribute equally to the two types.
This cannot be a significant source of confusion at 6\mic, where the NLR dust
contribution is small, or when using small aperture observation that exclude most of the NLR emission.
Is it possible that all these effects conspire to smear out the differences between type-I and type-II AGNs?
There were several attempts to blame the unexpected constant ratio of L(MIR)/L(2--10 keV) on the method used to estimate L(2--10 keV)
\citep{Brightman2011b,LaMassa2011,Mayo2013} and to obtain absorption-corrected X-ray luminosities that take into account more complex clumpy tori,
including partial covering in some directions. However, it is difficult to see how all this can  explain the above luminosity ratio..

\subsection{Imaging Interferometry}

Perhaps the most important development regarding AGN unification is the significant improvements in long baseline interferometry
and the ability to resolve the central structure on a milli-arcsecond scale.
The most significant results were obtained by three instruments:
1. MID-infrared interferometric Instrument (MIDI): The ESO interferometer working in the 8-13\mic\ atmospheric window with projected baselines ranging from about 30 to 130m
\cite[e.g.][]{Jaffe2004,Kishimoto2011a,Burtscher2013,Honig2013}. This instrument provided the largest sample.
2. The Keck K-band interferometer (KI) \cite[see][]{Swain2003,Pott2010,Kishimoto2011b} with a projected baseline of 85m.
3. The ESO Astronomical Multi-BEam combineR  \cite[AMBER][]{Weigelt2012}.
Interferometric measurements allow to measure and model the size and shape of the IR emitting components on a very small spatial scale of less than 1 pc.
The attainable resolution depends on the flux distribution in the source, the wavelength and baseline of the experiment, the redshift and the source 
coordinates (which determine the coverage of the uv-plane).
The results are often quoted in units of $R_{1/2}$, which is the wavelength dependent radius containing half of the emitted 
flux. The MIDI latest results, summarized in \cite{Burtscher2013} and illustrated in Figure \ref{fig:leonard_MIDI}, include 23 sources
with 8--13\mic\ dimensions of 1-100 pc. The KI and AMBER samples are smaller but they provide valuable information about shorter wavelengths and hotter dust.

%
%

The large scatter in the measured $R_{1/2}$ at MIR wavelengths is very different from the very tight correlation between size and luminosity indicated
by dust RM that probes regions closer to the inner torus boundary. 
Part of this must be related to the fact that in many cases, there are two distinct components in the resolved N-band interferometry. Translating
this to a specific size is obviously model dependent and subjected to large uncertainties. 
This indicates a rather complex relationship between radius and dust temperature in the torus.
Another surprise is the very similar size of type-I and type-II objects of the same luminosity (simple
tori predict that type-II tori will look larger). Also, 
The measured K-band $R_{1/2}$ is larger, by 0.2--0.4 dex, than the measured K-band RM size  \citep{Kishimoto2011b}. This is likely
related to the different definitions of the two radii since $R_{1/2}$ is a flux weighted radius and $R_{in}$ (or $R_{sub}$) a response weighted radius.

\begin{figure}[ht!]
\includegraphics[angle=0,width=0.6\textwidth]{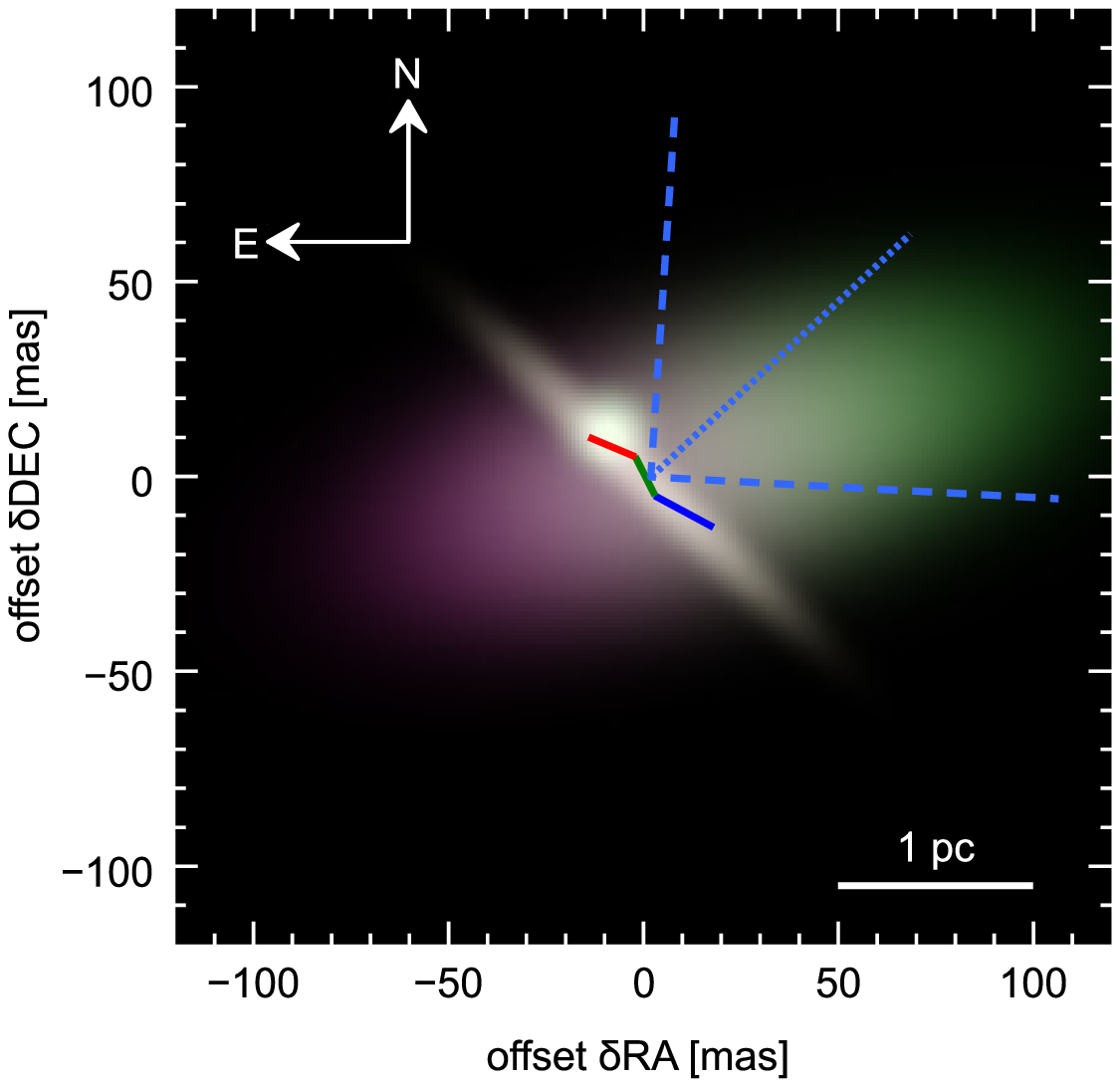}
\caption{
A MIDI-based model-image of the Circinus galaxy showing the location of the maser disk (red-green-blue line), central thin MIR emitting structure (white), and a larger thick 
elongated structure with colors marking central wavelengths: red - 13\mic, green - 10.5\mic\ and blue - 8\mic.
The thick red-green-blue elongated structure emits about 80\% of the MIR radiation. Its orientation is very different from the expected orientation of a ``standard'' torus.
The boundaries of the ionization cone (only one side cone in this source) are shown in dashed blue lines
(adopted from \cite{Tristram2014}, courtesy of Konrad Tristram).
}
\label{fig:circinus_MIDI}
\end{figure}

In a very small number of well observed nearby sources, the interferometry allows to obtain  ``model images'' that
reveal the rough structure, elongation, and inclination of the central dusty structure. As of 2014, there are four sources with 
this kind of information: NGC\,1068 (type-II), Circinus (type-II), NGC\,424 (type-I) and NGC\,3783 (type-I).
Of the four, the spatial information obtained for the Circinus galaxy, a type-II AGN with a graphite sublimation radius of about 0.01 pc, is perhaps the best. 
The emissivity maps of this source,  in three MIR bands, were modeled by \cite{Tristram2014}
and their combination is shown in Figure \ref{fig:circinus_MIDI}. The figure shows also the boundaries of the ionization cone based on HST imaging of the
\OIII\ line.
The map shows two distinct components. The first is a thin (axis ration of 1:6 or less) $\sim 1$ pc disk-like structure, 
emitting about 20\% of the MIR flux, with orientation which agrees
with the orientation of the smaller ($0.1-0.4$ pc) warped maser disk observed in this source \citep{Greenhill2003}. The
second is a much larger and thicker elongated structure, roughly perpendicular
to the first component.
This elliptical shape component, elongated roughly in the polar direction, emits about 80\% of the 8--13\mic\ flux 
with a clear wavelength gradient reflecting the dust 
temperatures in the various parts. None of this is in agreement with the shape and orientation of a simple central torus.

A second exciting result is obtained by combining MIDI (N-band) and AMBER (K-band) measurements of the type-I AGN NGC\,3783.
The combined model-image indicates different elongations of the K and N-band isophotes. The most likely elongation of the K-band model-image (which is still somewhat uncertain)
is in a direction parallel to the expected plane of the torus. The N-band model-image is of much larger dimensions and is elongated almost perpendicular to the K-band model-image,
 along the expected polar axis. 
  This geometry was interpreted  by \cite{Honig2013} as a combination of an inner small torus, where the dust is very hot, and a dusty polar outflow which, because of its distance, contains
  cooler dust. This is somewhat similar to the Circinus model mentioned earlier (where K-band information is not available). 
Interestingly, earlier ground-based AO-assisted observations pointed out polar emission in a different source \cite[NGC\,1068, see][]{Bock2000}.

  While complicated radiative transfer effects may result in the unusual model-images presented here \cite[see][]{Schartmann2005}, the
interferometric results for Circinus and NGC\,3783, and indications for similar structure in two other sources, suggest a 
possible change of  paradigm. A central equatorial, small covering factor torus, may still be needed in order to explain the hot dust emission, but much of the cooler $\sim 300$ K
  dust may well reside in an additional component which has different geometry, size and optical depth and, perhaps, different dynamics (i,e, a wind component). 
  A realistic model for this new structure must combine several different elements and take into account the fact that the hotter inner torus must also emit some of the 
   observed MIR flux (e.g. 20\% in the case of Circinus). 
   It should also answer several outstanding questions about the clear pass to the ionization cone, indicated by strong unreddened narrow lines,
 if the flow is optically thick, or the
lack of silicate emission in the Circinus spectrum if it is optically thin.
 One must also dismiss covering factor estimates that apply only to simple axisymmetric tori.
  In particular, the assumption of \Cf(total)=\Cf(hot dust)+\Cf(warm dust) may not work in this type of geometry. 
  Finally, it is interesting to note that several hydrodynamics models
  suggest dusty polar outflows  \citep{Wada2012} which may go part of the way towards in explaining
  the observations.

\section{ADDITIONAL AND ALTERNATIVE OBSCURERS}
\label{sec:additional_obscurers}

\subsection{Galactic-scale Obscuration}

Galaxy scale obscuration is common in many AGNs and numerous individual cases studied by HST  and 
ground based observations
have been documented since the 1980s
\cite[e.g.][]{Lawrence1982,Antonucci1993,Malkan1998,Prieto2014}. 
Well known cases are some of the objects detected by optical spectroscopy and classified here as type-Ii AGNs. In 
several past publications they were referred to as narrow line X-ray galaxies (NLXGs). 
Some of the objects show kpc scale dust lanes, others might be affected by dust in
the torus. The fraction of such objects might have been underestimated since many real type-IIs \citep{Merloni2014},
especially those with low quality spectroscopic observations,  may belong to this group. 
As argued in \S7, this group is a major source of uncertainty when coming to estimate the relative fractions of type-I and type-II AGNs

Galactic obscuration is more common in low luminosity type-II AGNs.
High spatial resolution observations of such objects reveal kpc scale dusty filaments that seem to be connected with 
much smaller dusty features close to the nucleus. This  
led \cite{Prieto2014} and others to propose that a central torus is not required for AGN unification at low luminosity.
Large scale LOS absorption has also been found in several Compton thick AGNs studied with \spitzer\ \citep{Goulding2012}.
In this sample, AGNs in face-on galaxies show little  9.7\mic\  silicate absorption, as predicted by clumpy torus models,
 and AGNs in nearly edge-on galaxies, or in mergers, 
show much stronger silicate absorption indicating dust in the galaxy.
This adds to previous similar results obtained for Compton thin type-II AGNs \citep{Deo2007,Deo2009}.
 In some cases, galactic extinction is so large that it can completely obscure the central AGN at optical wavelengths \citep{Goulding2009}.
The connection, if any, to the heavily reddened AGNs mentioned earlier, is not yet clear.

Supplementing the detailed investigation of individual sources, are
lower spatial resolution studies of many thousands of AGNs like \cite{Lagos2011}. 
This study finds that spiral and elliptical type-I hosts are strongly skewed towards face-on galaxies, while 
type-II hosts are skewed towards edge-on orientations.
The paper also shows that Galactic dust absorption by itself 
cannot explain the absence of broad emission lines in type-II sources. The inclination of the host galaxy strengthens the idea that
the angular momentum of the material which feeds the black hole retains a memory
of its origin in the galaxy, i.e., statistically speaking, galaxy disks, central tori and inner accretion disks, are generally aligned.
This issue, that was investigated in numerous papers, is still unresolved. For example, 
\cite{Kinney2000} used information from radio jet position angles and claimed to find no correlation between jet direction 
(i.e. inner accretion disk orientation) and the disk of the host galaxy.

\subsection{X-ray Obscuration by Gas in the BLR}
\label{sec:xray_BLR}

The simple idea of a single torus-like obscurer suffers from various limitations, especially when it comes to very large column densities.
For example, the observed optical reddening is much smaller than the value obtained by translating the measured X-ray column 
to foreground extinction assuming galactic dust-to-gas ratio \cite[e.g.][]{Maiolino2001,Risaliti2002,Goulding2012}.
There are also X-ray variations that indicate large changes in the obscuring column on  very short time scales (a few days or less), 
which is not consistent with the 
idea that the obscuring material is outside the sublimation radius.
This raises the possibility that obscuration by dust-free gas inside the BLR affects the X-ray absorption and hence also the AGN classification.

The typical dimensions of BLR clouds are estimated to be $10^{13-14}$ cm and their typical column density is of order
$10^{23-24}$ cm$^{-2}$. This is roughly the size of the central X-ray source in low-to-intermediate luminosity AGNs (a few $R_g$). 
For an X-ray continuum source $\sim 10^{14}$ cm across, and a cloud velocity of order 3000 \kms, the occultation time is of order
4 days. Can such clouds occult the central source giving the impression of a variable X-ray source? Recent observations and modeling suggest that this is indeed the 
case in almost every object where the observations are of long enough duration to discover such ``eclipses''.

There is now evidence from very detailed studies of a small number of type-I and type-II AGNs, for eclipses by large column density blobs
\citep{Risaliti2007,Risaliti2011}.
In the best studied case, NGC\,1365, the X-ray source underwent a total eclipse in less than 48 hours and then emerged
within the same time \citep{Risaliti2007}. Detailed modeling of this and similar events, by \cite{Maiolino2010}, provided information about
the likely location, velocity and size of the occulting gas, all consistent with being part of the BLR and inside the sublimation radius.
\cite{Maiolino2010} speculated that the occulting clouds
might have developed cometary tails, which are manifested as a gradual decrease in the column density as clouds emerged from the eclipse.
Different time dependent X-ray absorption events from inside the sublimation radius are described in \cite{Turner2009}.

There are more studies that are not as detailed but include a larger number of AGNs and occultation events.
The work of \cite{Markowitz2014} is based on long term monitoring of about a dozen AGNs of both types. These objects were monitored continuously by the 
Rossi X-rat timing explorer (RXTE) and the total duration is of order 10 years.
Given the known BH mass in these sources, one can work out the expected crossing time of dusty and dust-free
blobs of certain properties at different locations. 
The results suggest that many of the clumps are, indeed, inside the sublimation radius with properties that are consistent with being BLR clouds. Other blobs
are larger and move slower with locations that are consistent with being inside the dusty torus. Interestingly,
in several of the cases the clump location, assuming Keplerian orbits,
seems to be close to the maximum allowed dimension of such tori. \cite{Markowitz2014} presented also an analysis of the non-variable X-ray absorbing columns
and claim they are in conflict with the predicted dimensions, column density and number of clumps in clumpy torus models.
An additional study, by \cite{Torricelli2014}, includes more sources but shorter duration monitoring sensitive to events that last a few days at most. 
It presents numerous events associated with changing in hardness ratio of the X-ray continuum, all consistent with being caused by transiting blobs
inside the BLR. The observations are not detailed enough to allow a robust determination of the column density of the absorbers.

The X-ray observations provide a beautiful illustration of the connection between dust-free and dusty clumps in one general region that contains the BLR, closer
in, and the torus, further away from the central BH. The earlier apparent disagreement between optical reddening and X-ray column density disappears
when taking into account the  combination of inner dust-free and outer dusty absorbers. The observations also help to explain
the nature of some of the largest obscuring columns found
in AGNs.

\subsection{Tilted Disks}
Is an axisymmetric torus-like structure the only type of a central obscurer that can explain the various aspects of AGN unification?
Can other structures help to resolve   several of the known disagreements between simple torus models and observations?
An interesting scenario is based on the idea that a central disk can replace the torus 
\citep{Lawrence2010}. This ad-hoc phenomenological model is based on the idea that the angular momentum of the external 
inflowing gas is initially mis-aligned with the axis of the central accretion disk. The mis-alignment is kept down to
very small distances, of order one pc, and the inflow takes the form of an inclined, large dusty geometrically thin disk,
which may also be warped (see \cite{Caproni2006} for general mechanisms resulting in warped disks).
The large disk presents a wide solid angle to the inner disk and part of the UV-optical radiation is absorbed and re-emitted in
the IR. This simple geometry was used by \cite{Lawrence2010} to predict the covering factor distribution which was later studied by
\cite{Roseboom2013} using \WISE\ data of a 
large type-I sample. There seem to be a reasonable agreement between model and observations but the predicted
is similar to a normal distribution, which is expected also in other scenarios.

The main virtue of the tilted disk idea is the assumed mis-alignment between different accretion episodes onto the BH. 
This is expected in accretion events whose final stage take place in a region which is 2--3 orders 
of magnitude smaller than the region where the flow originates.
Unfortunately, the exact geometry has not been shown
to be consistent with realistic hydrodynamical simulations, the location of the sublimation surface has not been
calculated and compared with RM results, and there is no comparison with IR interferometry, IR SED and 
high quality polarization measurements.

\section{IONIZATION CONES}
\label{sec:ionization_cones}

An axisymmetric central obscurer with a central opening results in an anisotropic  
two-sided conical shaped ionizing radiation pattern (``bicone''). 
  Early observations \citep{Pogge1988,Tadhunter1989,Mulchaey1996} identified such cones in a number of nearby systems 
 and used them to deduce the torus inclination angles. 
The most detailed early studies are those of the archetype source, NGC\,1068, and the  
most detailed mapping utilized
the strong \OIII\ line (the ``\oiii\ bicone'').
  The more recent studies include larger samples, more detailed imagining in  the light of more emission lines, 
  and additional kinematic information. The mapped emission lines now include 
low ionization species like \oi,
  intermediate ionization ions like \oiii\ and \oiv, and very high ionization ions like \sivi.
High resolution images obtained with the Chandra X-ray observatory (\chandra) allow mapping such bicones in individual X-ray
lines like \ovii.

  The most detailed observations of today are  based on HST imaging, HST long slit spectroscopy, and ground based integral field unit (IFU) studies
  \cite[e.g.][and references therein]{Fischer2013,Mueller2011,Riffel2013}.
  The number of sources mapped in detail is more than 50 and those with reliable kinematic information more than 20. 
  The highest redshift is about  0.05. The observations are detailed enough to allow photoionization modeling of the gas in
   different parts of the NLR including, in rare cases, following the metallicity
  gradient across the source \cite[e.g.][]{Dopita2014}.

The ability to distinguish low and high ionization regions across a spatially resolved NLR, provides evidence
for ionization outside of the nominal \oiii\ bicone. In several well studied cases, there is evidence of ``filtering''  
   of the ionizing radiation in the parts producing the strongest low ionization lines by 
clumpy, ionized absorbers located tens of parsecs or less from the nucleus, consistent with the torus dimension.
A nice visual evidence is provided in NGC\,4151, where HST imaging shows that the \OIII/\OII\ line ratio 
is lower near the edges of the bicone compared with its axis \citep{Kraemer2008}.
 Such filtering was suspected in earlier works but without the 
 information about the location of the filtering material \citep{Alexander2000}. 
  
Line radiation from ionization cones provide ways to improve the estimate of the
bolometric luminosity of type-II AGNs where
the non-stellar continuum is not directly observed. In particular, the \OIII\ line was found to be a good
luminosity indicator because the line is strong and not affected much by stellar absorption \citep{Heckman2014}. 
This method
is not without difficulties. The line is not a very good luminosity indicator in low ionization AGNs like LINERs \citep{Netzer2009a} and it is now known that
L(\OIII)/\LAGN\ is changing with \LAGN\ \citep{Stern2012}. Line reddening is an additional uncertainty..
In particular, reddening by dust outside the NLR depends on the source geometry and torus inclination to the LOS and is,
therefore, different for type-I and type-II sources (line attenuation by dust inside the NLR clouds does not depend on these factors).
Indeed, observations show that the emission pattern
of some of the lines is more isotropic than in others 
which leads to the suggestion that \OIV\ is the best 
luminosity indicator because MIR reddening is much lower 
\citep{Kraemer2011,Weaver2010,Diamond-Stanic2009,Dicken2014}. 
In fact, I(\OIII)/I(\OIV) is typically lower by a factor of $\sim 2$ in type-II sources probably due to inclination-dependent reddening. 
Such a situation would naturally 
occur if the dust causing the extinction is just outside the \oiii\ bicone and the bicone axis is closer to the plane of the sky.

 Soft X-ray mapping of several NLRs  supplemented by \Chandra\ and \XMM\ high resolution spectroscopy, clearly indicate X-ray line emission, 
mostly \OVII, that generally overlaps with the \oiii\ bicone \citep{Sako2000,Ogle2000,Kinkhabwala2002,Bianchi2006,WangJ2011}. 
A spectacular example, showing X-ray emission up to about 7 kpc, in Mrk\,573, is given in \cite{Paggi2012}. 
The LOS attenuation of X-ray lines depends only on column density and metallicity, and not on the dust content in the gas. Therefore, observations
of X-ray lines provide an additional method to measure this column.
\cite{Kraemer2011} analyzed the \OVII/\OIV\ line ratio in a sample of about 15 sources and claimed an attenuation of the X-ray line by a column density which 
is consistent with the derived amount of reddening assuming ISM dust/gas ratio. While not
 a direct evidence, this is consistent with the idea of a dusty ionized gas outside and further away from the \oiii\ bicone.
Interestingly, X-ray emission is detected also in regions perpendicular to the optical bicone. This was interpreted as leakage through
a clumpy, relatively low column density absorber, which may be the torus.
\cite{Kreimeyer2013} used optical line ratios, to compare the flux emerging from the opening of the cone 
to the flux leaking through the torus. According to this work, the leakage amounts to 30--50\% of the radiation emitted by the central source.
This fraction is very  high and confirmation in other sources is required.

The kinematic studies provide information about the motion of the line emitting gas inside the cones \citep[e.g.][]{Mueller2011,Fischer2014}.
\cite{Fischer2013} presented multi-slit observations of 17 AGNs, 12 of which are type-IIs, with clear signatures of biconical outflows.  
The velocity maps allow  fits to simple kinematic models which are detailed enough
 to infer the inclination of the NLR to the LOS. 
In general, the angle of inclination is larger in type-II AGNs consistent with the view that these sources are seen edge on.
Interestingly, there is no correlation in this small sample between the NLR inclination angle and the orientation of the disks of the host galaxy
(see both supporting and contradicting results in earlier sections; this issue is not yet settled). 

\begin{figure}[ht!]
\includegraphics[angle=0,width=0.6\textwidth]{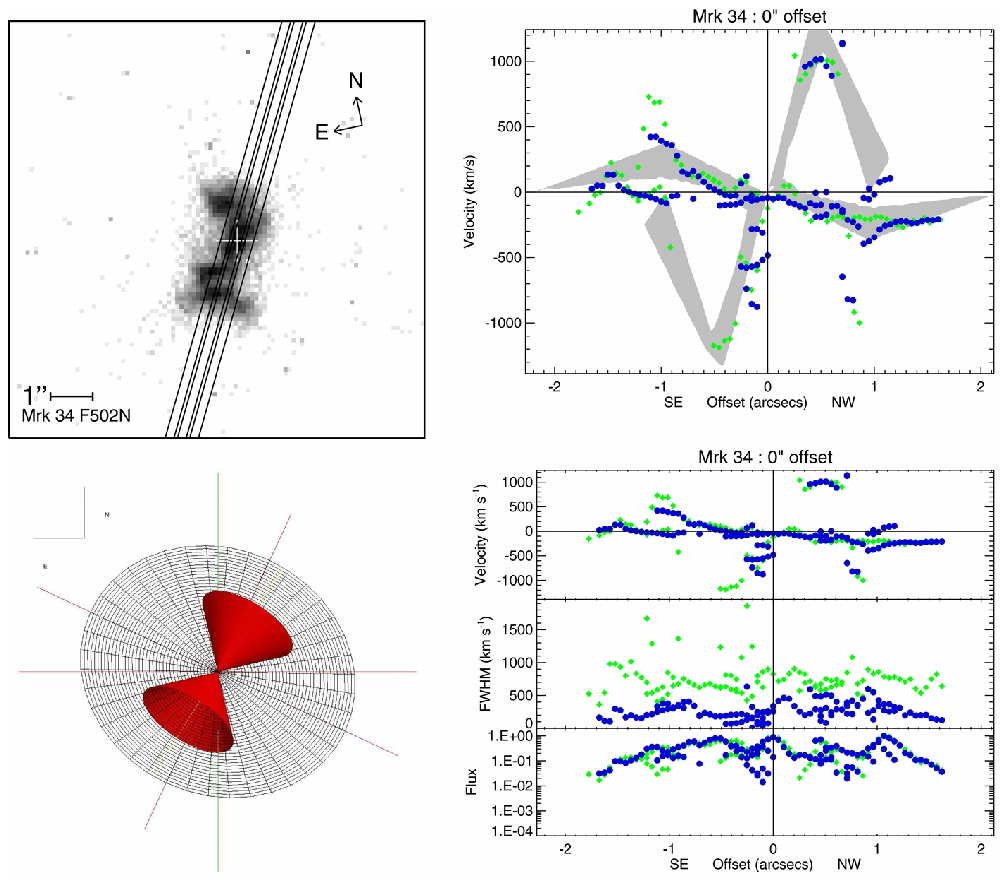}
\caption{
Kinematic mapping of the type-II source Mrk\,34 from \cite{Fischer2013} showing 
slit positions (top left), velocity, flux and FWHM of \OIII\ in different positions (right panels), and the 
resulted bicone geometry (bottom left). The green and and blue points represent data obtained with the G430L and G430M gratings, respectively.
}
\label{fig:mrk34_bicone}
\end{figure}

Kinematical modeling of ionization cones 
by \cite{Fischer2013} provides new invaluable information about ionization cones.
It reveals a unique pattern of gas outflow with $v\propto r$ out to
a certain radius, about 630$L_{46}^{1/2}$\,pc, and slowing down to zero beyond this radius.
 The fitted models are good enough to constrain the
 bicone inclination to the LOS and, in some cases, its opening angle.
Another interesting finding is a strong 
increase of X-ray column density with increasing angular distance from the bicone axis, regardless of
whether the column is neutral (type-IIs) or ionized (type-Is). For the type-I sources, where the X-ray absorbing gas is highly ionized (a warm absorber)
it suggests a connection between the inner walls of the torus and the origin of the X-ray ionized material.
The study  also suggests a correlation between mid-IR color and polar angle (see \S\ref{sec:torus_observations}).
For a handful of type-I sources, there is a hint for increasing FWHM(\hb) with polar angle. If verified in
larger samples, it would indicate a rotational component for the BLR gas.
Unfortunately, the results cannot be applied to the entire population partly because only one in three sources
had good enough data to be modeled in this way, and partly because in some of the sources the model is based on a single slit observation.
One good case is shown in Fig.~\ref{fig:mrk34_bicone}.

The \cite{Mueller2011} work supports some but not all the findings in \cite{Fischer2013}. 
This work provides imaging and kinematics of the coronal line region, the innermost part of the NLR where the strongest lines are
of very high ionization, e.g. \SiVI\ with ionization potential of 167 eV. In this region, the gas is probably dust-free.
The study of seven sources shows a complex dynamics of the gas including a rotating disk and an outflow bicone.
There are agreements and disagreements between the two works.
This is illustrated in the case of NGC\,3783, a well known type-I source with a large column of LOS ionized gas and 
interferometric data suggesting dusty polar outflow (see \S\ref{sec:torus_observations}).
\cite{Fischer2013} modeled this source using HST imaging and a single slit position. They found 
small inclination of 15 degrees to the LOS, as in a type-I AGN.
The IFU measurements of \cite{Mueller2011} resulted in a kinematic
map based on \SiVI\ and other coronal lines in a region very close to the base of the ionization cone. It suggests a large inclination, close to edge-on view, 
with the cone opening cutting into the plane of the host galaxy. If correct it means that the broad emission lines are seen through the (leaky?) torus.

In conclusion, the study of ionization cones provides valuable information about the central obscurer in AGNs. The detailed
mapping in the light of numerous lines confirms the general bicone shape and shows well defined borders between ionized and neutral gas.
It also reveals the kinematics of the gas and the overall inclination to the LOS.
The inferred models are more concrete and better justified experimentally than the rather vague information 
about the torus provided by 
fitting NIR-MIR SEDs.

\section{THE TORUS GALAXY CONNECTION}
\label{sec:torus_galaxy_connection}

General scenarios linking BH feeding and SF activity have been described in numerous papers and reviewed recently in this
journal by \cite{Heckman2014}. These concern global galaxy properties and apply to SF activity in the central kpc and beyond. 
They do not address the more direct connection between the two types of activity very much closer to the central BH.
What is known about the central 10--100 pc of nearby AGNs? What is the stellar population in this region,
the atomic and molecular gas content and the SF activity?
Is there evidence for a large central disk and is it connected to the torus?

Detailed observations of the central regions in nearby AGNs shed light on some of these questions.
Adaptive optics (AO) assisted observations, like in \cite{Davies2007}, can probe regions that are 10--30 pc in size.
 As of 2014, there are about 15 sources observed at this resolution.
Another work, by \cite{Esquej2014}, uses seeing limited N-band spectroscopy in 29 sources and probes a region of $\sim 65 $ pc in size.
The aim is to investigate SF near the nucleus using the 11.3\mic\ 
PAH emission feature.
About 45\% of the objects in the sample show this feature and stacking the spectra of the remaining sources suggests its  presence
in many others. There are several other works using seeing limited optical observations, with a typical spatial resolution of 
100--300 pc \cite[e.g.][]{Riffel2010,Riffel2011}. 
Important questions related to the conditions in a region which is slightly larger than the
size of the torus, or the sphere of influence of the central BH, are related to the age of the stellar population and the SF rate (SFR).
As explained below, there is no consensus regarding these issues.

Standard SF indicators are PAH features, optical absorption features and hydrogen recombination lines. 
The recombination lines are the best direct indicators for on-going SF (which is defined here as the presence of  O-type stars).
According to \cite{Davies2007} and \cite{Davies2014}, the equivalent widths of the $Br_{\gamma}$ lines in the 15 AGNs studied by
this group,  practically rule out on-going SF and indicate post SF population. On the other hand, \cite{Esquej2014} claim 
to detect SF activity based on the detection  of PAH emission features. 
This apparent contradiction can be understood by noting that PAH emission is not necessarily related to the presence of O-type stars. In fact,
there are indications that PAH emission is more sensitive to B stars than O stars, i.e. to an environment
that formed stars in the past 100 Myr or so \citep{Peeters2004,Diaz2010}. Moreover, the calibration of L(PAH) versus L(SFR) 
used by \cite{Esquej2014} to derive the SFR is based on work by \cite{Diamond-Stanic2012} and \cite{Rieke2009} who used templates of SF galaxies.
This uses SFRs and PAH luminosities 
obtained through large apertures that occasionally include the entire galaxy, and is equivalent to integrating over  $\sim 100$ Myr of SF history.

Returning to AGNs, the observed PAH emission close to the center is entirely consistent with the idea that the presently observed BH accretion
episode is  a few Myr old and BH accretion might have started after all the O-type stars in the region ended their life. In fact, the
onset of AGN activity might have terminated all SF in the immediate vicinity of the BH which is usually
categorized as ``AGN feedback'' \citep{Davies2012,Krisp2011,Hailey2012}.
Additional support for the post SF environment, and occasionally an even older stellar population very close to the
central BH, is obtained from other studies of the 100--300 pc central region 
\cite[][]{Riffel2010,Riffel2011,Storchi2012,Schnorr2014}.

Some observations have sufficient spatial and spectral resolution to derive the mass and age of the stars in
the inner 50 pc. 
For the few cases this was done \cite[e.g.][]{Davies2007}, the velocity field is dominated by random 
motion typical of a spherical system and the stellar mass is of order $10^8$ \msun. This mass is
an order of magnitude larger than the mass of the central BH in question. 
These observations set the stage for the stellar-wind model of \cite{Schartmann2010} discussed in \S\ref{sec:torus_models}.
The idea of BH feeding by stellar wind material has also been explored in connection with LINERs 
\cite[][and references therein]{Heckman2014}. The argument in this case is not based on spatially resolved observations
but rather on the extremely low \Ledd\ in LINERs. It is not at all clear whether the two independent scenarios have
anything in common.

Studies of molecular gas in the vicinity of the torus do not yet have the required spatial resolution to resolve the torus, 
or even its close surrounding, and hence were given low priority in this review. The coming ALMA observations will, 
no doubt, change this situation
\cite[see review by][]{Maiolino2008}. It is already clear that high concentrations of molecular gas on $\sim 100$ pc scales are 
seen in many AGNs with column densities that are large enough to confuse the pure torus absorption picture. 
There are clear indications for outflows that are possibly associated with the
direction of the bicone, but no indication of inflow. Some of these studies 
are detailed in \cite{Hicks2009}, \cite{Garcia2014}, \cite{Combes2013} and \cite{Aalto2014} where references to older observations can be found.

None of the above observations and theoretical considerations can be applied, by simple scaling, to higher BH mass, more luminous AGNs.
First, present day observations do not have the spatial resolution needed to investigate regions of this
size in such objects.
Second, feeding $10^{8-9}$ \msun\ BHs in high ionization AGNs by stellar winds would require a much larger stellar cluster 
and may involve very different conditions.
Direct mass inflow from the host is likely to play the central role in such cases.
Finally we note recent results from \cite{Merloni2014} that compared BH accretion rate and torus properties with total stellar mass 
and SFR of the host galaxy.
They indicate that the kpc scale stars and gas do not seem to have a direct causal connection with the BH and its immediate vicinity.
A similar lack of correlation at low redshift is shown in \cite{Koss2011}.

\section{UNIFICATION OR EVOLUTION?}
\label{sec:unification_or_evolution}

\subsection{Luminosity and Redshift Dependencies}

As explained, there are two categories of methods used to estimate \Cf.
The first category include all methods for measuring the covering factor of individual objects
(see \S\ref{sec:torus_observations}).
They include the \Ldust/\LAGN\ method, the EW(\Ka) method, and the method based on the scattered X-ray continuum. 
The  \Ldust/\LAGN\ method is the
one resulting in the largest number of measured \Cf.
The second category includes methods that apply to the  mean covering factor of the population. This is obtained by classifying a source
as either type-I or type-II AGN, and using the resulting N(type-II,L)/N(type-I,L) as an estimate of the mean value of \Cf.
One such method is based on the measured X-ray absorbing column and the AGN type is determined relative to a fiducial absorbing column that separates the groups.
\cite[e.g.][]{Ueda2003,Gilli2007,Hasinger2008,Brightman2012,Malizia2012}.
Another is based on counting the number of broad and narrow emission line AGNs in optical and IR selected samples.
Ideally, all methods should agree but this is certainly not the case \cite[e.g.][]{Lawrence2010,Mayo2013,Lusso2013,Lusso2014}).
The reasons range from the poorly understood physics of the torus, to various selection effects,
uncertainties in estimating \LAGN, intrinsic reddening, and more. For example, the \Ldust/\LAGN\ method is based on dust emission from the torus and
has little to do with BLR properties. However, X-ray absorption by gas in the BLR can increase the total measured X-ray column 
thus decreasing  \Ldust\ since the ionizing radiation absorbed by the dust-free gas never interacts with the dusty torus.
Here I focus on the large sample approach, in particular on X-ray based studies.

The hard X-ray (E$> 2$ keV) spectrum of most AGNs can be described by a simple power-law,  
$N_{ph} \propto E^{-\Gamma}$, 
where $N_{ph}$ is the number of emitted photons per unit energy and time. 
A remarkable feature is the very small scatter in the power-law slope, $\Gamma_{\rm 2-10 keV} = 1.8 \pm 0.2$.
This allows us to estimate the obscuring column density even in poor X-ray observations
provided we are looking through a {\it single} absorber.
All that is required is a measure of the hardness ratio, e.g. F(0.5--2 keV)/F(2--10 keV), and the redshift. The resulting accuracy in measured column is quite remarkable
 ($\sim 0.3$ dex), with even higher accuracy in cases of high photon flux that allow better modeling of the intrinsic continuum.
The hydrogen column densities derived in this way range from very thin (about $10^{21}$ cm$^{-2}$) to Compton thick
($> 1.25 \times 10^{24}$ cm$^{-2}$).
Some  of the results obtained so far are hampered by
selection effects. In particular, type-II sources are more
likely to drop below the flux limit of the samples because of the increased absorption. This is less noticeable, but still present, in samples
selected by their very high energy flux \citep{Malizia2012}.
The hardness ratio method can fail, sometimes dramatically, if the absorption is due to two or more LOS components, such as a combination 
of BLR clouds and a dusty torus gas (\S 4.2)

A very detailed study of X-ray obscuration is \cite{Merloni2014}. This work is 
based on observations of the two-degree XMM-COSMOS field and includes 1310 sources of all types selected on the basis of their
absorption corrected 2--10 keV luminosity in the redshift range 0.3--3.5. The rest-frame 2--10 keV flux limit is $2 \times 10^{-15}$ \ergs~cm$^{-2}$. 
The study supersedes earlier work of this type
\cite[e.g.][]{Ueda2003,Gilli2007,Hasinger2008,Malizia2012} that used data from other fields and instruments. It supplements a comprehensive study
of type-I sources in the same field by \cite{Lusso2013}. The spectral
properties of the sources are described in \cite{Brusa2010} and \cite{Mainieri2011}. 
The complete redshift information is based on spectroscopy and photometry
and is  of crucial importance since it allows to use a simple K-correction and thus circumvent
the well known bias of flux limited samples that tend to pick up more obscured sources at higher redshifts. 
A potential complication is a large number of Compton thick sources that are unaccounted
for. This number is still uncertain and depends on the survey, energy range used, and the spectroscopic analysis.
It is generally estimated to be $\sim 20$\% at low redshift \citep{Burlon2011} and $\sim 40$\% at high redshift \citep[e.g.][]{Brightman2012}. In the Cosmos sample which is not very deep, the numbers are smaller.
There are also issues regarding the best way to correct the X-ray luminosity for absorption, in particular the simplistic way based only
or hardness ratio with its well known limitations. 
  
An independent method to verify the estimated numbers of obscured and unobscured AGNs is based on  
modeling the spectrum of the cosmic X-ray background \cite[e.g.][]{Comastri1995,Ueda2003}. 
I will not go into the details of this method which was reviewed and discussed in numerous papers \cite[e.g.][]{Gilli2007}.

An important question is whether type-I and type-II sources classified by their optical properties can be distinguished, unambiguously, by their X-ray obscuring column? 
The answer is definitely no although type-II are, on the average, more obscured. \cite{Merloni2014} introduced a dividing column density 
of $N_H \simeq 10^{21.5}$ cm$^{-2}$ to separate the two types. They found about 30\% optically classified type-I sources with obscuring columns larger 
than this value and about 30\% of optically classified type-IIs with obscuring 
columns smaller than this column. In fact, many, perhaps most of the objects in these two groups may have little to do with obscuration by the torus.
The sources in the latter group are the real type-II AGNs discussed in earlier sections, and at least some of the objects in the first 
group are type Is with an X-ray source occulted by gas in the BLR or outflowing ionized gas in the opening of the torus..
These two sub-groups were not fully appreciated by the old unification scheme  and are definitely influencing the
N(type-I)/N(type-II) statistics.

Many of the real type-II AGNs in the local universe are of very low luminosity and \Ledd\ \citep{Brightman2011b,Marinucci2012}.
Those in the XMM-COSMOS sample, at redshift of $\sim 0.3$, are more luminous and can be considered
intermediate luminosity AGNs.
An important clue to the nature of these sources is the presence or absence of a central obscurer. This can be tested by examining the 
L(MIR)/L(2--10 keV) relationship
discussed in \S\ref{sec:torus_observations}. The real type-II objects in the \cite{Merloni2014} sample do not differ in this respect 
from other type-IIs.
There is also direct evidence from \spitzer\ spectroscopy
that the MIR SEDs of these sources are similar (but not identical) to the SEDs of type-II sources showing hidden broad emission lines \citep{Tommasin2010}. 
It seems that many real type-II AGNs do not have enough high density gas to result in a detectable BLR 
but have central tori. This raises an interesting question about the 
relationship between the central 
source luminosity, the inner dimension of the torus, and the
presence or absence of broad emission line gas. This question is discussed in \S\ref{sec:receding} below.

Returning to the type-I sources with X-ray absorption, we note that 
for a Galactic dust-to-gas ratio, the above dividing column corresponds to $A_V \sim 1.7$ mag. If the obscurer is part of the dusty torus, 
then the intrinsic optical-UV spectrum of 30\% of all type-I objects 
are expected to be heavily reddened. This is definitely not the case.
As noted earlier, the
obscuring gas may be part of the BLR in which case it is dust-free. Alternatively, the absorbing gas 
is highly ionized outflowing, dust-free  material (``warm absorber'' which is known to be common in AGNs \citep{Turner2009}.
The location of this gas maybe well outside the BLR, including in the opening of the torus.

\begin{figure}[ht!]
\includegraphics[angle=0,width=0.7\textwidth]{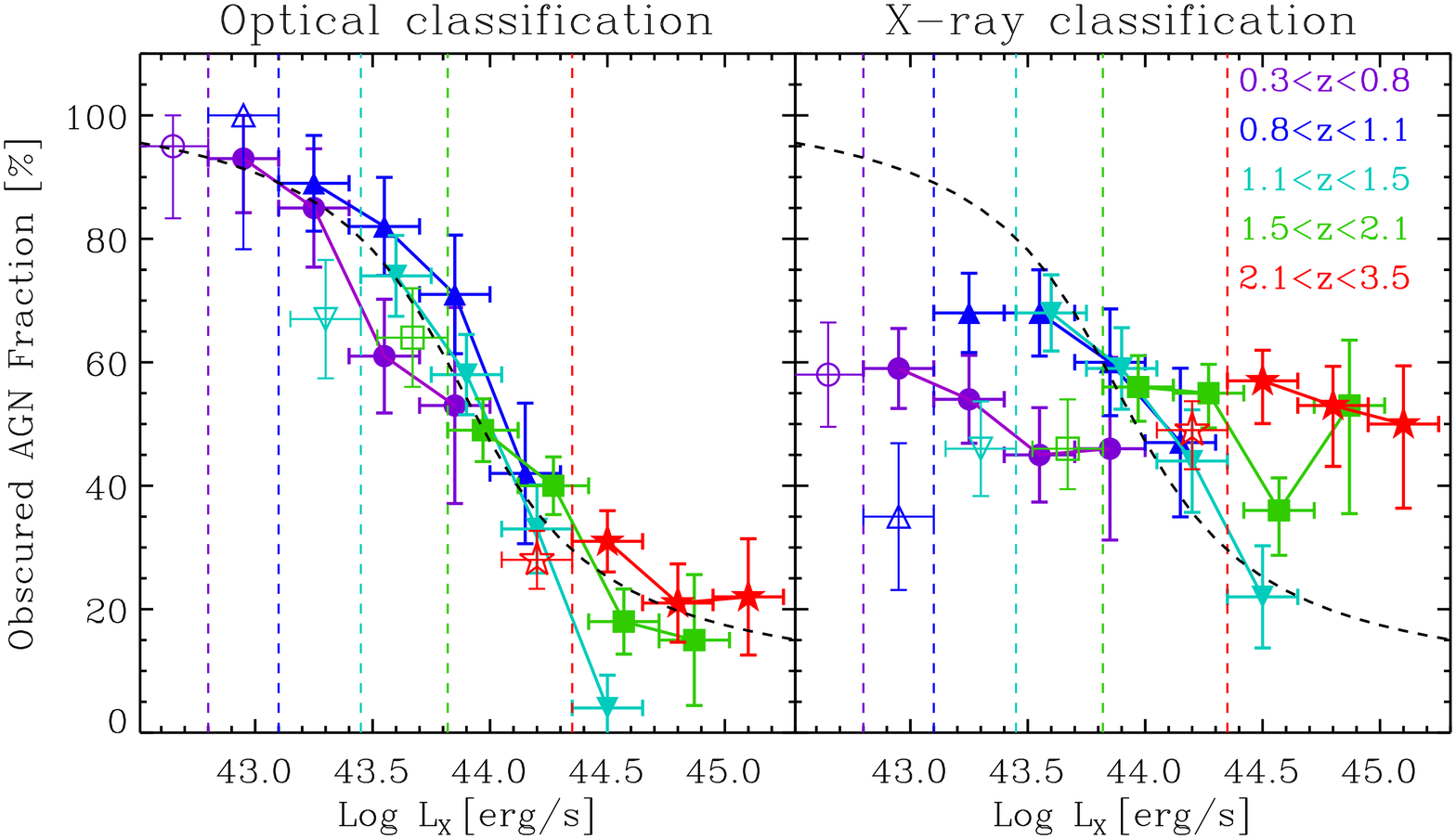}
 \caption{The fraction of obscured and unobscured AGNs, at various redshifts and L(2--10 keV), from \cite{Merloni2014}.
The left panel shows  the fractions obtained by using optical classification. The right panel 
uses a classification based on column densities derived from X-ray observations with a dividing column between the groups of $10^{21.5}$ cm$^{-2}$.
 }
 \label{fig:merloni2014}
\end{figure}

Figure~\ref{fig:merloni2014} shows the fraction of obscured AGNs defined in two different ways as a function of L(2--10 keV). 
The dependence on X-ray luminosity of the two groups is very different, showing a steep decrease in obscuring fraction if classified by optical 
properties, and a very flat and weak
dependence if X-ray classification is used. Clearly, the large number of real type-II AGNs 
is the main reason for the difference between the two methods at low luminosity. 
This suggests a new, perhaps more intuitive  definition of AGN types based only on the presence or absence of a central LOS obscurer. 
According to this classification, real type-IIs belong in the class of type-Is (unobscured AGNs). In this case, the relative fraction of the two 
groups up to redshift of about 1.5, is about 1:1 according to both the X-ray and optical methods.

The \cite{Merloni2014} sample doses not include low redshift very low luminosity AGNs and a couple of notes about this end of the distribution
are in order. First, the IR-selected
sample of \cite{Brightman2011b} shows that
the ratio N(type-II)/N(type-I) peaks at around L(2--10 keV)$\sim 10^{42}$ \ergs\ and drops to lower values {\it below} this luminosity. 
Second, the discrepancies illustrated in Figure~\ref{fig:merloni2014} are present in various other studies, including those  
 that are based on the measurements of individual covering factors
(that are, occasionally, in poor agreement between themselves). The sources of confusion have been discussed earlier in detail. A major one is,
again, the confusion between host galaxy and torus obscuration which is related to the large 
group of type-Ii sources.

What is the redshift evolution of obscured and unobscured AGN? 
Observationally, this is very challenging because of the very small number of observed sources at high redshift. Earlier works like \cite{Hasinger2008} claim an increase
in the fraction of obscured sources with redshift. This was questioned by \cite{Gilli2010} on grounds of a lack of an adequate K-correction.
The \cite{Merloni2014} work, which is perhaps the most complete (a statement that will likely be challenged by some X-ray astronomers), is inconclusive 
about this issue regarding low and intermediate luminosity sources. The results for the highest luminosity AGNs are more significant with indications for increase in the
fraction of obscured sources with redshift amounting to a factor of about 2 between redshifts of 1 and 3. This finding is supported by
several studies of AGNs in the \chandra\ Deep Field South (CDFS) \citep{Brightman2012,Vito2013}.

\subsection{Receding and Approaching Tori and Disappearing BLRs}
\label{sec:receding}

There were several attempts to explain the dependence of covering factor on source luminosity based on the torus geometry.
The idea of a ``receding torus'', first suggested by \cite{Lawrence1991}, involves a very simple torus whose height, $h$, 
is independent of the inner radius $R_{in}$. Since 
$R_{in} \propto $\LAGN$^{1/2}$, 
the covering factor
is predicted to be proportional to \LAGN$^{-1/2}$. This ad hoc model  makes no attempt to justify the constant height torus. 
It also does not explain the decreasing covering factors deduced from counting X-ray AGNs with different obscuring columns.

Extensive studies based on large AGN samples, some of which have already been mentioned, have been used to test the receding torus idea.
None of these reproduced, accurately, the predicted dependence of
\Cf\ on \LAGN. There were several suggestions to modify the original idea. For example,
\cite{Simpson2005} measured the fraction of type-I sources by using \OIII\ luminosities in SDSS AGNs and constructing LFs for the two types of AGNs. 
The differences between the two LFs were explained to be due to changes in the torus covering factor with luminosity 
and were claimed to be removed if 
$h \propto {\rm L_{AGN}^{0.23}} $. 
Unfortunately, the work does not take into account the fact that the conversion between L(\OIII)
and \LAGN\ is luminosity dependent \citep{Stern2012}, a fact which was not known at the time. \citet{Ricci2013} used the observed EW(\Ka) to suggest a 
somewhat different dependence. 
These attempts, and others, are highly simplified and should not be viewed 
as more than fits of simple analytic functions to the observations.   
The increased inner radius with luminosity is, indeed, a solid and confirmed prediction of all torus models.
However, other parameters, like \Cf\ and $h$,
 are not well defined and there is no simple theoretical reason to prefer a certain geometry over others.

The idea of a luminosity dependent sublimation radius has also been used to argue that below a certain \LAGN, the torus is likely to change its properties,
and even disappear. This idea deserves the name ``approaching tori''. Several of its aspects were
described in \cite{Laor2003}, \cite{Elitzur2006} and  \cite{Elitzur2009}.
 \cite{Laor2003} suggested to  combine the sublimation radius as the outer boundary of the BLR, with the observed relationship 
between 
BH mass and \LAGN. This gives an estimate of the minimal broad line velocity  (taken here as the full-width at half maximum, FWHM, of
the broad \hb\ line), which is the Keplerian velocity at $R_{in}$.  
Using current RM results one finds,
\begin{equation}
FWHM(H_{beta}) \simeq  1700 M_8^{1/2} L_{46}^{-1/4} {\rm km\,s^{-1} },
\label{eq:FWHM_hb}
\end{equation}
where, again, $L_{46}$=\LAGN/10$^{46}$ erg~s$^{-1}$ and $M_8=M_{BH}/10^8$ \msun.
For $M_8=1$ and $L_{46}<10^{-4}$, FWHM(\hb)$>17,000$ km~s$^{-1}$, a line
width that is practically undetectable in low luminosity AGNs whose continuum near the \hb\ line is dominated by stellar light.
Additional considerations show that the line width may be related to \Ledd\ rather than \LAGN.
\cite{Laor2003} further speculated that there is a natural limit to the maximum allowed velocity in the BLR and clouds in this region may not 
survive close to the BH,
not just become undetectable.
Low luminosity objects that appear to have only narrow high ionization lines will be classified
as real type-II AGNs and low ionization line sources will be classified as type-II LINERs. This results in the increasing fraction
of (apparent) type-IIs at low luminosities. The argument does not affect the torus covering factor. It
applies mostly to extremely low luminosity AGNs in the local universe 
but not to the real type-IIs in the \cite{Merloni2014} sample where \LAGN$>10^{43}$ \ergs.

Disappearing BLRs have been discussed by \cite{Nicastro2000} who suggested that gas outflow from the surface of the disk, outside the region dominated by
radiation pressure, can lead to disappearing BLRs at very low \Ledd.
\cite{Elitzur2009} followed the ideas developed in \cite{Elitzur2006} and considered the clumpy dust-free BLR, and the clumpy
dusty torus, as one entity made of clouds lifted along magnetic field lines from the surface of the disk. A limiting radial column density of $\sim 10^{22}$ cm$^{-2}$,
was assumed to be a requirement for producing observable broad emission lines from the clumps. The work shows that at low mass accretion rate, corresponding to
\LAGN$ \sim 10^{40}$, BLR clouds, and large dusty clumps, no longer exist. 
This luminosity, which varies with the black-hole mass, and the 
corresponding \Ledd, are in the right ball park
for many LINERs but is several orders of magnitude smaller than the typical luminosity 
of many real type-II AGNs  \citep{Brightman2011b,Marinucci2012,Merloni2014}. Despite the discrepancy, the general idea can perhaps be extended to include 
more luminous sources with no detectable broad emission lines.

\subsection{Mergers Evolution and Unification}

Detailed morphological studies of host galaxies of AGNs, up to redshift $\sim2$, reveal that about $\sim 80$\% of them 
do not show signs of interaction \cite[e.g.][]{Cisternas2011,Schawinski2012}. Such studies depend on signal-to-noise, spatial resolution, and
most importantly redshift. They should be interpreted as strong evidence against major mergers and perhaps not so strong evidence regarding the final
stages of a minor merger where the mass ratio of the two hosts is large.
The finding, which contradicts earlier claims about the
association of most AGN activity with galaxy mergers, is evidence for the importance of secular processes
in transporting cold gas from the disk to the vicinity of the BH. 
The studies show also that a large fraction
of the most luminous AGNs at high redshift are associated with mergers. Indeed, merger induced nuclear
activity is likely to be more common at very high redshift, when the density of the universe was higher and much of cold
gas was not yet converted to stars. Can simple unification based on relatively simple shaped tori, and well ordered motion of
cold gas, be applied to the complex geometry of such mergers? How early in the history of an AGN triggered by a major merger 
can such an ordered central structure form?
Can merger driven AGNs be classified like the secularly evolving systems if their BLR is obscured by galactic scale dust 
\citep[e.g.][]{Imanishi2010},
and what are the implications to the NIR-MIR SEDs of such systems?

Major mergers of gas rich systems in the early universe have been suggested to result in powerful dusty SF systems (dust-obscured galaxies, DOGs)
surrounding a buried AGN whose power is increasing with time
\citep[see][and references therein]{Narayanan2010}. As the AGN power increases,
the surrounding gas is blown away and an unobscured AGN is revealed. 
The most luminous AGNs during the era of peak AGN activity, at z=2--3, have been suggested to be objects of this type. The geometry of the obscuring dust in this case must
be very different from the simple tori assumed in non-merging galaxies.  
Unfortunately, numerical merger simulations do not yet have sufficient resolution to model the geometry of such structures and high resolution imaging is limited to nearby systems
(see however \cite{Schawinski2012}).
Thus, SED observations and model fitting are the best tools to investigate these scenarios, in particular in objects where both optical
and IR information is available \cite[e.g. SDSS and \wise\ see][]{Weedman2012,Vardanyan2014}. 

An empirical way to study such scenarios is to examine objects with very large
\Ldust/\LAGN\ and check whether this is caused by UV 
luminosity which is unusually low (UV extinction in a DOG) or MIR luminosity which is unusually high (large covering
 factor and a clear LOS to the center).
The work of \cite{Vardanyan2014} suggests that the latter is correct and shows that the objects with the smallest \Ldust/\LAGN\ are those with the largest L(UV).
Such studies show no evidence of the DOG-related scenario in z=2--3 AGNs
(see however \cite{Glikman2012} for evidence to the contrary). 

There are more detailed studies of merging systems at low and intermediate redshifts, especially
ultra luminous infrared galaxies (ULIRGs) hosting active BHs.  A low redshift sample of this type is the QUEST 
(Quasar/ULIRG Evolutionary Study) sample \cite[see][and references therein]{Veilleux2006,Veilleux2009a,Veilleux2009b}.
Some of the sources in the sample contain low luminosity AGNs, including LINERs, and are clearly SF dominated systems. Others contain
high luminosity AGNs, mostly PG Quasars..
HST imaging of QUEST sources indicate that the type of AGN depends on the
evolutionary stage of the host, and the BH activity is most likely triggered by the merger.
In particular, there is evidence that a merger progresses from type-II at the earlier stages,
to type-I closer to its end. Type-I AGNs become more and more common among fully-coalesced merger remnants with
elliptical-like hosts \citep{Veilleux2006}.
They also become more common with increasing \Ldust. This may or may not be related to the general
trend of decreasing obscuration with increasing L(2--10 keV) discussed for non-merging systems.
Many of the QUEST sources have been observed, spectroscopically,
by \spitzer. Interestingly, their comparison with sources that show no indication of mergers \citep{Mor2009,Mor2012}
shows very similar SEDs and a similar
trend of decreasing covering factor with \LAGN.

Can major mergers at high redshift produce galactic-scale obscuration that results in torus like SED and absorption?
Can heating by the AGN result in cold (30--50 K) dust emission that mimics SF regions?
There are no clear observational answers  to these questions but recent calculations by \cite{Schneider2014} are already giving some clues.

A complex geometry that can give rise to both MIR and FIR emission, is a ``torus within a torus'' or ``an obscurer within an obscurer''.
In this scenario, a large covering factor obscurer, just outside the dust sublimation radius, results in intense
NIR-MIR emission. This obscurer is clumpy and has a simple axisymmetric geometry. Part of the AGN radiation escapes through its ionization cones
and part leaks out through the clumps.
A much larger obscurer, with an undefined geometry that does not block the central dust-free cone, is present at a distance
of a few kpc. The covering factor of the inner obscurer is \Cf(in) and the one of the outer obscurer \Cf(out).
It is easy to think of several scenarios that result in  different relative L(FIR)/L(MIR)/\LAGN, depending on how much radiation
of the central source leaks through the small obscurer near the center.
It is also possible that some of the MIR radiation of the central obscurer is absorbed by the further dust, if its optical depth is large enough.
A certain variant of this complex geometry was modeled by 
\cite{Schneider2014} who found that such a geometry with \Cf(in)$\sim 0.7$, gives a reasonable fit to the observed SED of a specific z=6.4 AGN.
In that particular model, about half of the FIR luminosity is AGN-related and half SF-related.

 An obvious observational test would be to obtain polarimetry maps of such systems since this is a most effective way to check whether the 
distribution of the position angles is centrosymmetric (i.e. scattering from a central point)  or not \cite[see][]{Kishimoto1999}.

In summary, it is hard to imagine that major mergers, especially in the early universe, can lead to the highly ordered geometry, and the
torus-like structure, suggested to explain the observations of lower luminosity, secularly evolving systems. Detailed SED studies of high redshift
mergers can help to resolve this issue.

\section{SUMMARY: PRESENT AND FUTURE UNIFICATION}

The unification scheme presented in this review is different in some ways, and similar in others, to the scheme presented
in the late 1980 that had three fundamental pillars: orientation, covering factor and luminosity. The field
is driven by courageous attempts to construct simple models for complex situations, and attempts to understand sever selection effects typical
of large samples.
The most important developments of the last decade or so can be summarized as follows:
\begin{enumerate}
\item
Most AGN tori are probably axisymmetric and clumpy. 
Unfortunately, todays phenomenological clumpy torus models leave much to be desired and their parameters are hard to constrain.
Hydrodynamical calculations that follow the inflow of gas from the galaxy, are more promising. They naturally produce a clumpy thick 
central structure and
demonstrate the importance of SN explosions, and radiation pressure feedback by the central AGN. The simulations do not yet include all 
relevant  processes and are generally limited to AGNs with small BHs.
\item
Disk-wind models of different types seem to be able to explain several fundamental properties of the central obscurer, 
 like the formation of dust-free and dusty clumps at roughly the desired locations.
So far, there is no clear preference for magnetic or radiation pressure driven winds and there is a general problem, in all models,
to form and maintain large column density clumps. 
\item
The central issues of torus covering factor and anisotropy are not yet fully understood. The general trend of decreasing \Cf\ with increasing \LAGN\ is 
confirmed but its details not well understood. There are substantial
disagreements between the various methods used to study this relationship, like the ones based on \Ldust/\LAGN, on EW(\ka), or
the division of the population into groups based on X-ray absorbing columns. 
The suggestion of a receding torus is the only idea driving this field. However, the idea is neither physically sound nor well
understood.
\item
Study of ionization cones and IR interferometry are two major developments in this field. Bicone observations 
provide the most reliable way to define
the torus boundaries and link  the emission line properties to the kinematic properties of the NLR gas.
Long baseline interferometry is the way of the future. It has already provided global intriguing results, like an unexpected run of
dust temperature with radius and model-images that indicate deviations from a simple torus-shaped obscurer.
They also stress the need to reconsider the role of MIR emitting polar outflows.
Future phase information will likely provide real images of central AGN tori.
\item
Part of the confusion in the present unification scheme is the presence of several sub-groups that may not belong in it in the first
place. 
In particular, ``real type-II AGNs'', that are not really understood, confuse several issues like the fraction of obscured and unobscured
low luminosity AGNs. Another group that shows a mixed bag of properties are type-Ii AGNs. Some of these sources are 
heavily extinguished by dust in the host galaxy, others are type-Is 
``caught'' during a time of minimum AGN activity.
\item
Unification does not seem to stand the test of BH evolution. In particular, merging systems likely behave in a different way from
AGNs in secularly evolving hosts. Based on todays observations, there is no reason to believe that accreting BHs in major mergers, or
merger remnants, partake in the simple, old  unification scheme.
\end{enumerate}
Returning to the central theme of unification, it is clear that the old scheme
requires several major revisions. The more important ones are the recognition that for a given luminosity
there is a large range of covering factors, the realization that X-ray absorption can be caused by two and possibly more components, and the
identification of several sub-groups of AGNs that do not belong in this scheme in the first place. A revised scheme would require better ways to
isolate and remove cases where obscuration has nothing to do with the nucleus, and improved X-ray techniques to clarify 
the importance of X-ray absorption by dust-free gas. 
Perhaps the most crucial aspect, which is observationally
challenging, is the separation of torus absorption from the presence of broad emission lines. 
Putting the absorption of the non-stellar continuum in the center of the scheme, and minimizing the use of broad emission lines as
a major characteristics, are good ways to proceed. 
%


\section{Disclosure statement}
The author is not aware of any affiliations, memberships, funding, or financial holdings that might be perceived as affecting the objectivity of this review.

\section{Acknowledgements}
This review benefitted from comments and suggestions by colleagues and friends who shared with me their broad knowledge and
deep understanding of this field. Special thanks go to:
Ski Antonucci,
Murray Brightman,
Leonard Burtscher,
Mike Crenshaw,
Ric Davies,
Michael Dopita,
Moshe Elitzur,
Martin Elvis,
Jacopo Fritz,
Josefa Masegosa Gallego,
Sebastian Hoenig,
Steven Kraemer,
Ari Laor,
Andy Lawrence,
Dieter Lutz,
Vincenzo Mainieri,
Roberto Maiolino,
Andrea Merloni,
Richard Mushotzky
Cristina Ramos-Almeida,
Guido Risaliti,
David Rosario,
Marc Schartmann,
Clive Tadhunter,
Konrad Tristram,
Jane Turner,
Mara Salvato,
Allan Schnorr Muller,
Marko Stalevski,
Sylvain Veilleux,
Keiichi Wada,
and 
Belinda Wilkes.



\end{document}